\documentclass{emulateapj}
\usepackage{apjfonts}
\usepackage{natbib}
\usepackage{graphicx}
\usepackage{amssymb,amsmath}
\usepackage{xcolor}
\definecolor{ao}{rgb}{1.0,0.03,0}
\usepackage[breaklinks,colorlinks,linkcolor=ao,citecolor=blue]{hyperref}
\shorttitle{Separating Gaussian CMB polarization signal}

\begin{document}

\title{Separating CMB Stokes Q and U polarization signals from Non-Gaussian Emissions}

\author{Ujjal Purkayastha\altaffilmark{1},  Rajib Saha\altaffilmark{1}}

\altaffiltext{1}{Physics Department, Indian Institute of Science 
Education and Research Bhopal,  Bhopal, M.P, 462066, India.} 

\begin{abstract}
In this paper we estimate diffuse foreground minimized Cosmic Microwave Background (CMB) 
Stokes Q and U polarization maps based upon the fundamental concept of Gaussian nature of 
CMB and strong non-Gaussian nature of astrophysical polarized foregrounds using WMAP nine
year published polarization maps. We excise regions of the sky that define position of 
the known point sources, regions that are strongly contaminated by either the detector 
noise or by the diffuse foregrounds or both, and then perform foreground minimizations 
over the surviving  sky regions that constitute approximately $50\%$ of the full sky area. 
We critically evaluate performance of foreground minimizations in several ways and show that our 
foreground minimization method removes significant foregrounds from input maps. 
The cleaned Stokes \{Q, U\} polarization maps have less EE and BB power from relevant sky region 
compared to WMAP foreground-reduced Stokes \{Q, U\} polarization maps at different multipole ranges. 
We validate our methodology  
by performing detailed Monte Carlo simulations. The main driving machinery of our method 
is an internal-linear-combination (ILC) approach, however, unlike simple variance minimization 
performed in usual ILC method, the core of the method being dependent on the theoretically 
well motivated concept of Gaussianity  of CMB polarization a direct connection is established 
between observations and models of inflation. Additionally, the  method, like the usual ILC 
method, is independent on modeling uncertainties of polarized foregrounds. It will be  useful 
to apply our method in future generation low-noise CMB polarization experiments.  
\end{abstract}

\maketitle

\section{Introduction}
\label{Intro}
Anisotropies of Cosmic Microwave Background (CMB) radiation are deemed to be the 
fundamental ingredients responsible for the structure formation and therefore for the very 
existence of the  Universe as we observe it and think of it today. Soon after detection 
of CMB \citep{Penzias1965} it was realized that the anisotropic universe can induce 
polarization in CMB with unique feature \citep{Rees1968,Negroponte1980,Lubin1983}. 
The anisotropies in CMB are expected to be weakly and linearly polarized \citep{cct_1995} 
due to Thompson scattering at the surface of last scattering  and due to imprints 
of small fluctuations in metric that could have been generated during the epoch of 
inflation \citep{GuthInflation1981, StarobinskyInflation1982, LindeInflation1983}. Although 
weak, Stokes Q and U  polarization anisotropies of CMB contain valuable information 
about various physical processes in a wide range of energy scales - starting from 
$10^{16}$ GeV or so, at the time of inflation  until today, when the energy scale of 
cosmological interest is only $\sim 0.2$ meV. To highlight the rich variety of problems 
that have been investigated in the literature in this context we mention that polarization 
anisotropy can be used to constrain  primordial gravitational wave \citep{Polnarev1985, 
CDS_1993, Frewin1994}. \cite{cct_1995} show that it is possible to infer about the 
primordial tensor anisotropies from the correlation of temperature and polarization of 
CMB. \cite{KKS1997} find a method that can probe long-wavelength gravitational wave and 
vector-mode perturbations of metric elements  using CMB polarization. \cite{Stark1981_a, 
Stark1981_b,Zaldarriaga1997} show that reionization of intergalactic medium induces 
new polarization in CMB. \cite{SpZal1997} argue that CMB polarization can serve as a 
direct test of inflation. The primordial gravitational wave as measured by CMB polarization 
can be used detect energy scale of inflation \citep{Knox2002}. The effect of gravitational 
weak lensing on the CMB polarization power spectrum and also on temperature-polarization 
cross power spectrum was studied  by \cite{ZalSel1998}. \cite{Hu2002} propose a method to 
reconstruct large scale mass distribution of the Universe using CMB polarization.                             
\cite{SelZal1999} show that it is possible to measure dark matter power spectrum
by using distortion of CMB weak lensing. \cite{NikhilDoug2005} proposes a method to constrain 
possible dark matter annihilation at the time of recombination using CMB polarization. 
\cite{Balaji2003} show that the parity odd Chern-Simons terms in the effective Lagrangian 
containing Maxwell fields can give rise to polarization axis rotation which in turn could probe
physics of very early universe. \cite{PlanckInflation2016} argue that CMB polarization signal 
is more powerful than temperature anisotropy signal to constrain any possible sharp features 
in the primordial scalar power spectrum. \cite{Bucher2001} showed that polarization observations 
can constrain primordial isocurvature modes by lifting its degeneracy with the cosmological 
parameters.  Detailed theoretical framework for the study of  CMB polarization 
have been developed by \cite{Kosowsky1996, Kosowsky1999, Cabella2004, Bucher2015}. 

Following the discussion above, it is clear that CMB polarization signal can be used to 
understand problems in Cosmology with  a diverse spectrum. Therefore, it becomes utmost 
important to device a method that estimates a clean map of the CMB polarization, which 
then can be reliably used to extract these information. In real life, however, estimating 
a cleaned CMB signal from observation  in microwave part of electromagnetic spectrum is 
a non-trivial task since the net polarization signal that can be measured by a  polarization 
sensitive detector placed on Earth, or  in space, contains both the weak primordial CMB 
signal along  with polarization signal due to various foreground emissions  that originate 
due to different astrophysical processes that take place inside our Galaxy.

An interesting property of Stokes Q and U polarization signals  of CMB generated due 
to primordial gravitational wave is that they follow  Gaussian distribution. Moreover
the polarization generated at the surface of last scattering due to Thompson scattering 
were directly proportional to the amplitude of quadrapolar temperature anisotropy of 
that time. Therefore, such polarizations signals again follow Gaussian distribution with 
a high accuracy.  The only assumption that we make in both the above cases is simple slow
roll models for inflation -- where any non Gaussianity generated are small \citep{Allen1987, 
Gangui1994, Gangui2002, Munshi1995, Acquaviva2003, Maldacena2003, Komatsu2003} - are 
valid descriptions. The anisotropies of the polarized foregrounds however follow highly 
non-Gaussian properties due to complex non-linear physics involved in their origin. Motivated 
by the very accurate  Gaussian nature of the cosmological Stokes Q and U polarization signals  
and highly non-Gaussian signature of polarized foregrounds, in this work, we seek to 
isolate  cleaned pictures of primordial CMB Stokes Q and U anisotropy signals from their 
observed mixture with foregrounds, based upon the Gaussian nature of the former and 
non-Gaussian nature of later. The method has been applied on WMAP temperature anisotropy 
data by \cite{Saha2011}. In this work, we extend the earlier work to the case of CMB Stokes Q 
and U polarization. A related approach in the context of CMB which is useful to estimate 
maximally independent components of a mixture is independent component analysis (ICA) 
(e.g., see \cite{ICA_Hyvarinen, ICA_2004}).   

   The basic driving machinery of our method is the usual internal-linear-combination (ILC) 
approach (e.g., see ~\cite{Bennett1992, Tegmark1996, Bennett2003, Tegmark2003, LILC_Eriksen04, Saha2006, 
Hinshaw_07, Saha2008, Gold2009, Kim_HILC_temp, Kim_HILC_Pol, Samal2010} for descriptions about ILC approach and 
applications thereof).  However, unlike the usual ILC method, wherein data-variance is minimized, in 
the current work, we minimize a suitably chosen statistic that quantifies the measure of 
non-Gaussian property intrinsic to a set of samples. As in \cite{Saha2011} we take sample 
kurtosis as the measure on non-Gaussianness of the samples  and minimize this quantity to 
estimate the cleaned CMB Stokes Q and U polarization signal following the ILC approach.  
Additionally, as discussed later in Section \ref{formalism}, our method relies on the blackbody  
nature of CMB spectrum \citep{Mather1990, Mather1994, Fixen1996} and non-blackbody nature of 
spectra of different foreground components. As a consequence of this, our final estimate of 
cleaned CMB signal also preserves any non-Gaussian CMB polarization signal along with the Gaussian 
CMB signal, when both types have a blackbody spectrum. Thus our method also preserves CMB weak 
lensing signal which possesses non-Gaussian properties along with the Gaussian signals due to 
primordial Gravitation wave and Thompson scattering at the surface of last scattering.

   The ILC approach has been applied  in the field of component separation in the context of CMB 
in many different forms. \cite{Delabrouille2009,Basak2012, Basak2013} implement a needlet space 
ILC approach on WMAP temperature and polarization maps. \cite{Remazeilles2011, Remazeilles2011a} 
generalize ILC method for non-CMB component estimations. \cite{SILC_2016_temp, SILC_2016_Pol} implement 
a scale dependent, directional ILC method on CMB temperature and polarization maps. \cite{Saha2016} 
apply the ILC approach following a perturbative technique on simulated Stokes Q observations of WMAP 
and Planck at low resolution to estimate variation of synchrotron spectral index over the sky 
positions and jointly estimate all foregrounds and CMB components. \cite{Saha2017} improve the 
usual iterative ILC  approach in harmonic space by nullifying a foreground leakage signal.  

   Among the non-ILC approaches we refer to \cite{Bunn1994} and \cite{Bouchet1999} for Weiner filter 
method,  \citep{Bennett2003, Hinshaw_07, Gold2011,  Gold2009} for template fitting methods,       
\cite{Gold2009,Gold2011} for Markov Chain Monte Carlo Method and \cite{Eriksen2007, Eriksen2008, 
Eriksen2008a, PlanckCol2016a, PlanckCol2016b} for Gibbs sampling approaches for component separation.

We organize our paper in the following way. We present  the problem that we wish to discuss in this 
work in  Section \ref{Problem}. We present our statistical variable to quantify  non-Gaussian 
properties and to minimize foreground contamination in Section \ref{NG-measure}. We then describe 
the basic formalism of this work in Section \ref{formalism}.  In Section \ref{inputs} we first discuss 
preparation of input maps of our method. We then briefly discuss foreground and noise 
properties of these input maps.  Based upon these discussions we identify regions of the sky where 
either foreground, or detector noise or both are dominant source of contamination. Using the 
knowledge of these regions, in Section \ref{mask} we describe the mask that removes regions with 
strong foreground or detector noise contamination. We  use this mask later in Section \ref{results} 
to reconstruct foreground minimized Stokes Q and U  maps. We then demonstrate 
the non-Gaussian nature of polarized foregrounds using the empirical foreground model proposed by the 
WMAP science team in Section \ref{ng-pol_fg}, both i) from the sky region defined by the mask obtained 
in Section \ref{mask} and ii) from the full sky. After demonstrating the non-Gaussian properties of 
polarized foregrounds in Section \ref{Callib} we then describe how the statistical variable to measure
non-Gaussian properties (e.g., see Section \ref{NG-measure})  varies with increase in contamination
strength of the non-Gaussian foregrounds.   In Section \ref{inpmap_kurt} we describe non-Gaussian 
properties of the input  maps due to foreground contamination. We discuss variation of the values 
of our  non-Gaussian measure with detector noise contamination but without any foreground in Section \ref{kurt_no_fg}.  
We discuss the foreground minimization methodology in Section \ref{method}. In Section \ref{results}
we discuss the results and critically evaluate the performance of the foreground minimization. In Section \ref{MonteCarlo}
we discuss the detailed Monte Carlo simulations of foreground minimizations and results thereof. 
In Section \ref{D&C} we finally discuss our work and conclude.   

\section{The problem}
\label{Problem}
The basic problem that we plan to discuss in this work is as follows. Let, we have observations 
of CMB Stokes Q and U parameters at different frequencies (and possibly at different instrument 
resolutions) over certain region of the sky. As discussed in Section \ref{Intro} slow-roll 
inflationary models with a single scalar field  predict that primordial anisotropies of CMB 
polarizations signal follow Gaussian distribution with a very high accuracy. On the other hand, 
astrophysical foregrounds strongly deviate from Gaussian nature due to complex non-linear processes 
involved in their emission process. We propose to provide a solution of  the question of separating 
CMB Stokes Q and U polarization signals from its observed  mixture with (non-Gaussian) foregrounds 
based upon the Gaussian nature of the former and non-Gaussian nature of later. We refer to 
Section \ref{Callib} for a discussion about non-Gaussian nature of foregrounds. 

\section{Measure of non-Gaussianity}
\label{NG-measure}  
Which statistical variable can be used as  the measure of non-Gaussianity?  
As in \cite{Saha2011} we take excess kurtosis as measure of non-Gaussian nature of 
a set of random values. The excess kurtosis for a set of $N$ random numbers  $R_k$ 
is given by
\begin{eqnarray}
\mathcal K = \frac{1}{N}\sum_{k=1}^N\frac{\left(R_k-R_0\right)^4}{\sigma^4} -3\,,
\label{excess_kurt}
\end{eqnarray}
where $R_0$ represents sample mean and $\sigma^2$ is the sample variance. Excess kurtosis 
for a Gaussian probability density function is zero. As described  later in Section \ref{Callib} 
the excess kurtosis for foreground contaminated Q and U signals tend to take positive values 
and increases quickly with the increase in the contamination levels. On the other hand, the 
excess kurtosis for a random Gaussian realization of CMB Stokes Q or U map is close to zero. 
This makes the excess kurtosis an efficient estimator for measuring non-Gaussian foreground 
contamination in CMB polarization maps. In subsequent part of this paper we use `sample kurtosis'
or simply `kurtosis' to actually represent the excess kurtosis as defined by Eqn. \ref{excess_kurt}.

\begin{figure*}
 \includegraphics[scale=0.25]{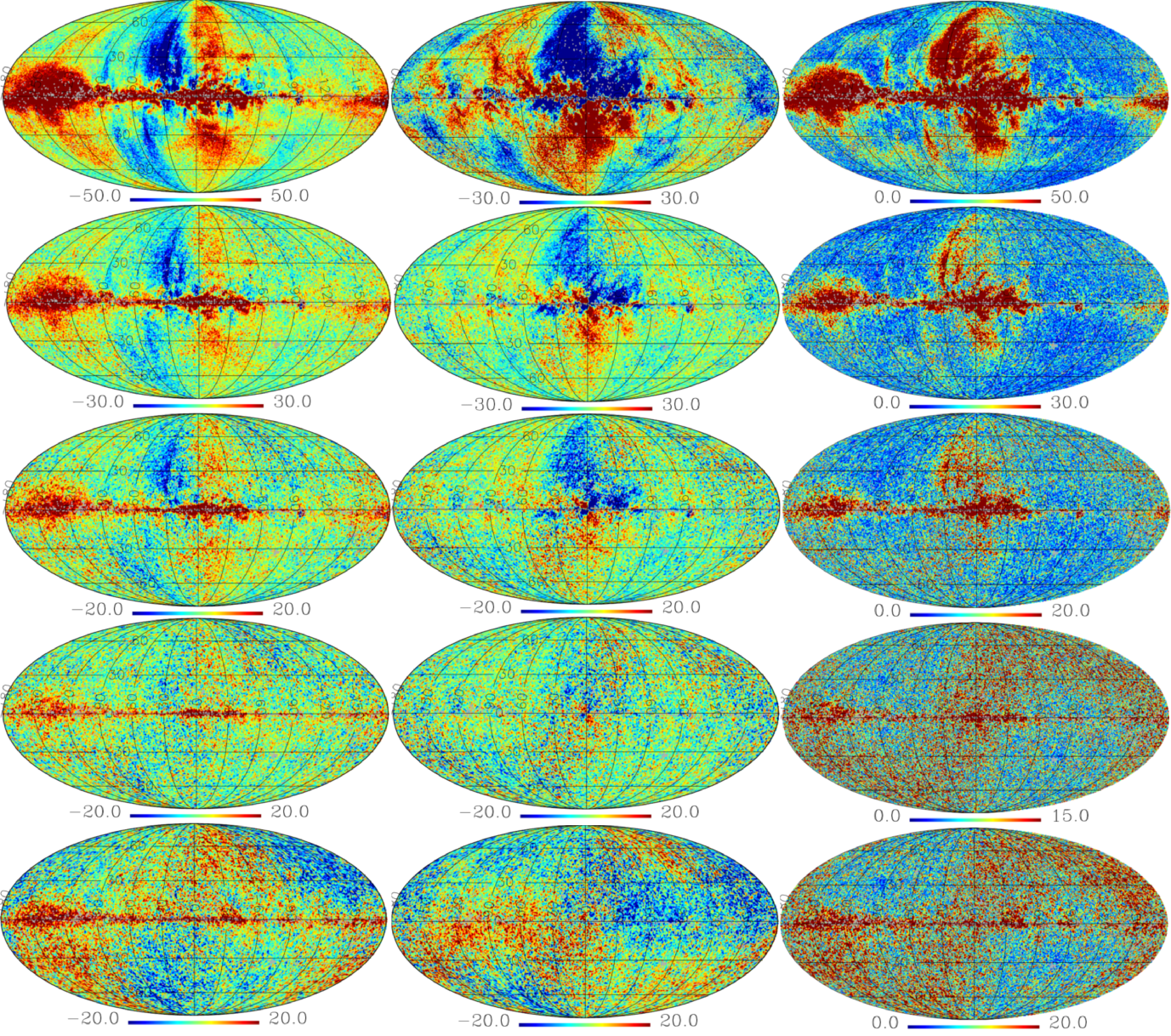}
 \caption{Left column: figures, from top to bottom, showing the Q polarization maps starting from WMAP K1
 to W band frequency maps, in the order of the increasing frequency of different bands. 
Middle column: same as left panel but for U polarization map. Right column: figures showing magnitude of   
polarization vector $P = \{Q^2 + U^2\}^{0.5}$ using different WMAP  frequency maps as shown in the first 
two columns. The low frequency Q and U maps are dominated by the diffused galactic contamination due to   
synchrotron component. The high frequency maps of V and W bands also contain detector noise as a dominant source 
of contamination apart from significant diffuse foreground contamination due to thermal dust component. 
The noisy nature of the polarization maps in the right panel are clearly visible for the  V and W band 
frequency maps. One of the important things that these maps show is that the WMAP low frequency maps are  
dominated by the foreground contamination whereas high frequency maps have both foregrounds and detector noise
as significant sources of contamination. All the maps of this figure have a common instrumental 
resolution of $1^\circ$.}
\label{fig:inp_maps}
\end{figure*}

\section{Basic Formalism}
\label{formalism} 

In this section we review the basic formalism (e.g., see \cite{Saha2011}) of estimating 
Gaussian Stokes Q and U polarized CMB components from the mixture with non-Gaussian 
foreground components. The basic driving tool of our  method is  an ILC method in pixel 
space to estimate foreground minimized CMB Stokes Q and U anisotropy maps. However the 
core of the method relies on a more fundamental concept of minimizing the non-Gaussianity
of the ILC map, rather than minimizing a relatively simpler concept of variance 
that is followed in the usual ILC methods. 

Let we have observations of CMB Stokes Q or U polarization over some region of the sky
at $n$ different frequency bands. Further, we assume that all these maps have the same instrumental beam 
and pixel resolutions \footnote{In general, and as is the case of WMAP, the observed maps at 
different frequency bands possess different instrumental resolutions, even if they have 
same pixel resolution. Here we assume that all such maps have already been converted 
to a suitably chosen and common instrumental resolution. If the frequency maps had different pixel 
resolutions as well, one could use the method described by \cite{Saha2017}.} and they have already 
been converted to thermodynamic ($\mu K$) temperature unit. If $M_i(p)$  represents Stokes Q or U 
polarization signal at pixel $p$ and at a frequency $\nu_i$ where $i \in \{1,2,3,...n\}$ then 
\begin{eqnarray}
M_i(p) = M^c(p) + M^f_i(p) + M^n_i(p) \,,
\label{freq_map}
\end{eqnarray}
where $M^c(p)$ represents the CMB polarization anisotropy signal at a pixel $p$.  CMB anisotropy 
is independent on frequency $\nu_i$ in thermodynamic temperature unit since frequency 
spectrum of CMB follows blackbody spectrum. $M^f_i(p)$ represents the net foreground contribution 
at frequency $\nu_i$ and at a pixel $p$. Finally, $M^n_i(p)$ represents the detector noise 
contribution at a frequency $\nu_i$ and at a pixel $p$. We note that, in Eqn. \ref{freq_map} we have 
omitted any explicit reference of beam or pixel window function. This is justified since we assume 
all frequency maps $M_i(p)$ have same beam (and also pixel) smoothing effects. In other words, each 
component of the right hand side of Eqn. \ref{freq_map} should be interpreted as appropriately 
smoothed versions of the actual sky components or detector noise. 

Using the frequency maps $M_i(p)$ the foreground minimized ILC map, $M^{\textrm C}(p)$, is defined 
as the linear superposition of all of them with certain amplitude terms  $w^i$ that depends upon the 
frequency band $\nu_i$
\begin{eqnarray}
M^{\textrm C}(p) = \sum_{i=1}^nw^iM_i(p)\,, 
\label{ILC}
\end{eqnarray} 
where the amplitudes  $w^1, w^2, ...., w^n$, defined as weight factors 
\footnote{Here we have omitted any dependence of weight factors with the pixel positions at any  
given frequency $\nu_i$. An important future project will be allowing weight
to vary with the sky positions, to take into account varying spectral properties of polarized foreground 
components, e.g., synchrotron and thermal dust components. Letting weights depend on the sky positions 
will also allow them to be varied with different regions of the probability density function of the net 
foreground emissions, which is expected to lead to better performance of our foreground removal.  We shall 
explore this problem in a future publication. In the current work we treat weights to be independent on 
sky positions for any given frequency.}. To reconstruct the CMB component which follows a blackbody spectrum we 
further impose a constraint on the weights that $w^1 +  w^2 +  .... +  w^n = 1$, i.e., the weights 
of all frequency bands sum to unity. Such a constraint leads to effectively $n-1$ independent number 
of weight factors.  The weight factors are obtained by minimizing `sample kurtosis' (e.g., see  
Section \ref{NG-measure} and Section \ref{Callib})  of the cleaned map.
 
If the cleaned map, $ M^{\textrm C}(p)$ contains $N$ pixels following Eqn \ref{excess_kurt} the 
sample kurtosis of the cleaned map is given by 
\begin{eqnarray}
\mathcal K^C = \frac{1}{N}\sum_{p=1}^N\frac{\left(M^{\textrm C}(p)-M_0\right)^4}{\sigma^4} -3\,,
\label{sample_kurt}
\end{eqnarray}
where $M_0$ denotes sample mean of the cleaned map and $\sigma$ represents standard deviation. 
Using Eqn \ref{ILC} and the condition on weights that they sum to unity one can show that 
\begin{eqnarray}
 M_0 = \sum_{i=1}^nM^i_0\,,
\label{mean}
\end{eqnarray}
where $M^i_0$ represents the mean temperature of input map at a frequency $\nu_i$.  Using Eqns \ref{ILC}
and~\ref{mean} along with the constraint equation satisfied by the weights, in Eqn. \ref{sample_kurt} we find 
\begin{eqnarray}
\mathcal K^C({\bf W}) =\left[\frac{N}{\left({\bf W M W}^T \right)^2} \sum_{p=1}^N\left( {\bf W M}_p{\bf  W}\right)^2\right] -3\,, 
\label{kurt_cmap}
\end{eqnarray}
where $\bf W$ is a $1\times n$ row-vector with $w^i$ as its $i^{th}$ element. ${\bf M}_p$ is 
an $n \times n$ symmetric  matrix for each pixel $p$ and is elements are  defined as, 
${\bf M}_{p(ii')} = \tilde M_i(p)\tilde M_{i'}(p)$ , where $\tilde M_i(p)$ represents the polarization (Q or U) 
signal at pixel $p$ for frequency $\nu_i$ after the mean signal corresponding to this frequency band
has been subtracted from the original pixel values. $\bf M$ in the denominator of Eqn. \ref{kurt_cmap}
is also an $n\times n$ matrix and is given by 
\begin{eqnarray}
{\bf M} = \sum_{p=1}^{N}{\bf M}_p\,.
\end{eqnarray}

\section{Input Maps}
\label{inputs}
\subsection{Data Preparation}
In this work we use WMAP nine year published Stokes Q and U polarization maps corresponding to 
different differencing assemblies (DA) as the primary input maps. A total of $10$ DA maps are provided by the WMAP 
science team at HEALPix\footnote{Hierarchical Equal Area Isolatitude Pixellization of Sphere, 
e.g., see \cite{Gorski2005}.} pixel resolution parameter $N_{\textrm side}  = 512$ at different
frequencies starting from $23$ GHz to $94$ GHz. Each of these DA map belong to a given frequency band.
A list of all the DA maps, their frequency of observations and corresponding frequency band are mentioned 
in Table \ref{ListMap}. Since each DA  map has  different instrumental 
resolution corresponding to different detector of WMAP satellite mission the DA maps are not directly 
usable {\it as is} in our method.  We first bring all the  DA maps to a common resolution corresponding 
to a Gaussian polarized instrumental response function of FWHM = $1^\circ$ at $N_{\textrm side} = 512$
by multiplying the spherical harmonic coefficients of each map by the ratio of beam window 
functions corresponding to $1^\circ$ Gaussian beam window and the native beam window function of 
the corresponding DA map\footnote{We use WMAP nine year published beam window functions corresponding 
to different DA maps in this work.}. Each of WMAP 23 and 33 GHz frequency bands (i.e., K1 and Ka1 bands) 
has only  single detector each. However, since each of WMAP 41 (Q), 61 (V) and  94 GHz (W) 
frequency bands has multiple DA maps  we average all the DA maps corresponding to each of these 
frequency bands to form a single map corresponding to each of these frequency bands. This 
results in a total of five frequency maps corresponding to all  K1, Ka1, Q, V and W bands of WMAP
for use as input in our work.  

\begin{table}
 \caption{List of WMAP maps used in this work}
 \label{ListMap}
 \begin{tabular}{lll}
  \hline
 Frequency &  Band Name & DA Maps  \\   
 (GHz)     &        &            \\   
 \hline
 23  & K1 &   K1  \\   
 33 &  Ka1 &  Ka1 \\
 41 &  Q   &  Q1, Q2 \\
 61 &  V   &  V1, V2\\ 
 94 &  W   &  W1, W2, W3, W4\\ 
  \hline
 \end{tabular}
\end{table}

\subsection{Foreground and noise properties}
\label{fg_nse_input}
We show these input frequency maps after masking of the position of known point sources 
in Fig. \ref{fig:inp_maps} in $\mu K$ thermodynamic temperature unit. The leftmost column of 
this figure shows Q Stokes maps,  in order of increasing frequency from top to bottom, corresponding to
five WMAP frequency bands. The middle column shows the same except for the U Stokes signal. The rightmost 
column shows the magnitude of polarization vector, $P = \{Q^2 + U^2\}^{0.5}$. Both the Q and U Stokes maps 
are dominantly contaminated by the galactic diffused contamination, both in the galactic plane and out of the 
galactic plane for K1, Ka1 and Q frequency bands. This is clearly seen from the plots of 
the first three rows of this figure. The foreground polarization in the V and W bands are  
more dominant in the galactic plane than the region outside the plane. The V and W band 
also contain significantly strong detector noise which can be easily inferred from the plots 
of the last two rows, specifically from the last two plots of the rightmost column. The transition 
from foreground dominated low frequency polarization maps to both foreground and detector noise 
dominated high frequency WMAP maps is clearly visible from the last column of this  figure. We 
note in passing that mean standard deviation from $1000$ simulations of  CMB Stokes Q and U 
polarization signal at the same $N_{\textrm side}$ parameter and $1^\circ$ resolution is merely 
$\sim 0.55$ $\mu K$. This indicates that all these frequency maps of this figure contain foreground 
and detector noise contaminations that are stronger than the expected level of 
primordial signal.       

\section{Masks}
\label{mask}
\begin{figure}
\includegraphics[angle=90,width=\columnwidth]{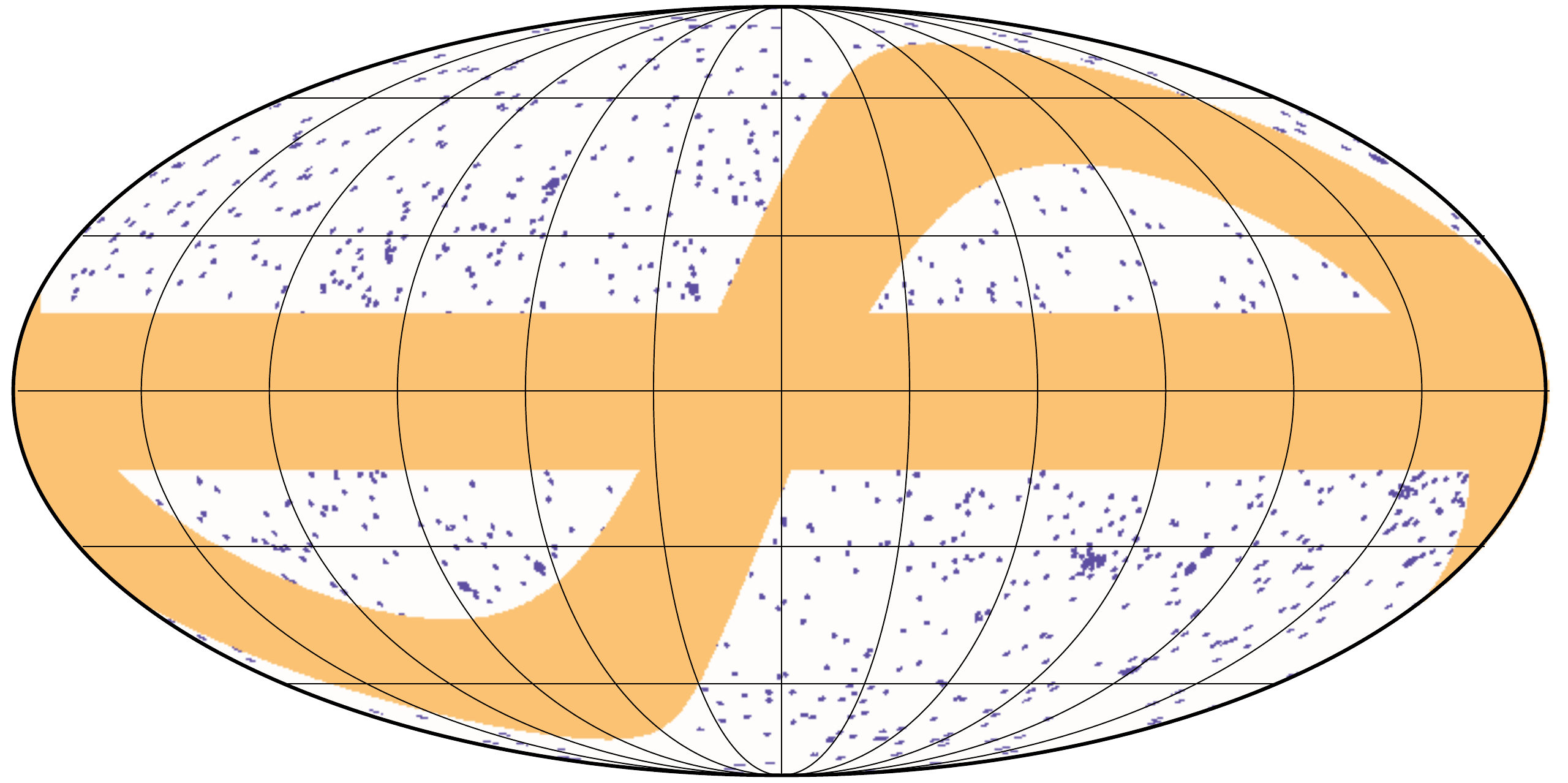}
\caption{The {\tt CMask} or composite mask  used in this work to define usable sky region for the polarized 
foreground removal. The golden yellow region shows the region defined by $\pm 15^{\circ}$ on both sides of 
the ecliptic plane and galactic plane and are excluded in this work. The  blue regions, which is also removed 
in this work,  represent the position of resolved point sources as given by the WMAP nine year point source 
mask outside the golden yellow region. The white region of this figure is used for foreground removal of  
this work.}
\label{mask.fig}
\end{figure}

As described in Section \ref{fg_nse_input} and as seen from Fig. \ref{fig:inp_maps} the low frequency 
polarization maps (K1, Ka1, Q frequency bands) of WMAP have strong foreground contaminations both 
along the galactic plane and away from the plane in both the hemispheres. 
The high frequency V and W band maps are more dominated by the detector noise along the ecliptic plane.
Presence of foreground contamination can be visibly seen along the galactic  plane in all the five frequency 
maps (e.g., see the third column of Fig. \ref{fig:inp_maps}).  Since  presence of any detector noise 
that is not negligible compared to the foregrounds  significantly reduces the  efficiency of foreground removal,
 we chose to remove regions on the sky that is heavily contaminated by the detector 
noise. For this purpose we make a mask in the ecliptic  coordinate system that excise $15^{\circ}$ on each side 
of the ecliptic plane. We find the spherical harmonic coefficients of this mask and rotate these coefficients 
to galactic coordinate system using the HEALPix provided facility {\tt alteralm}. We then convert the rotated 
harmonic coefficients to a new mask. The resulting mask contains some pixels with values differing from unity or zero. 
We finally convert this mask to a new mask by setting pixel values less than 0.5 to zero and rest to unity. 
The resulting mask is a binary map and excise all the regions defined by $\pm 15^{\circ}$ along the ecliptic plane 
in the galactic coordinate system. To reduce contaminations due to resolved point sources we also remove the 
position of point sources as defined by the WMAP nine year point source mask. Finally we  excise all pixels 
of this mask that lies within $\pm 15^\circ$ of the galactic plane. Our final mask is the product of the 
WMAP nine year point source mask, the $\pm 15^{\circ}$ mask along the ecliptic plane and the $\pm 15^{\circ}$ mask 
along the galactic plane. This final mask,  called  {\tt CMask} (or composite mask)  in our analysis, contains a 
total of $1611603$ surviving pixels at $N_{\textrm side} = 512$ implying a $51\%$ sky region being used 
in our analysis.  We show the {\tt CMask} in Fig. \ref{mask.fig}. The {\tt CMask} region described hereafter 
in this work refers to the sky region where {\tt CMask} takes the value unity. 

\section{Non Gaussian nature of polarized foregrounds from density analysis}
\label{ng-pol_fg}
\begin{figure*}
\includegraphics[scale=0.72]{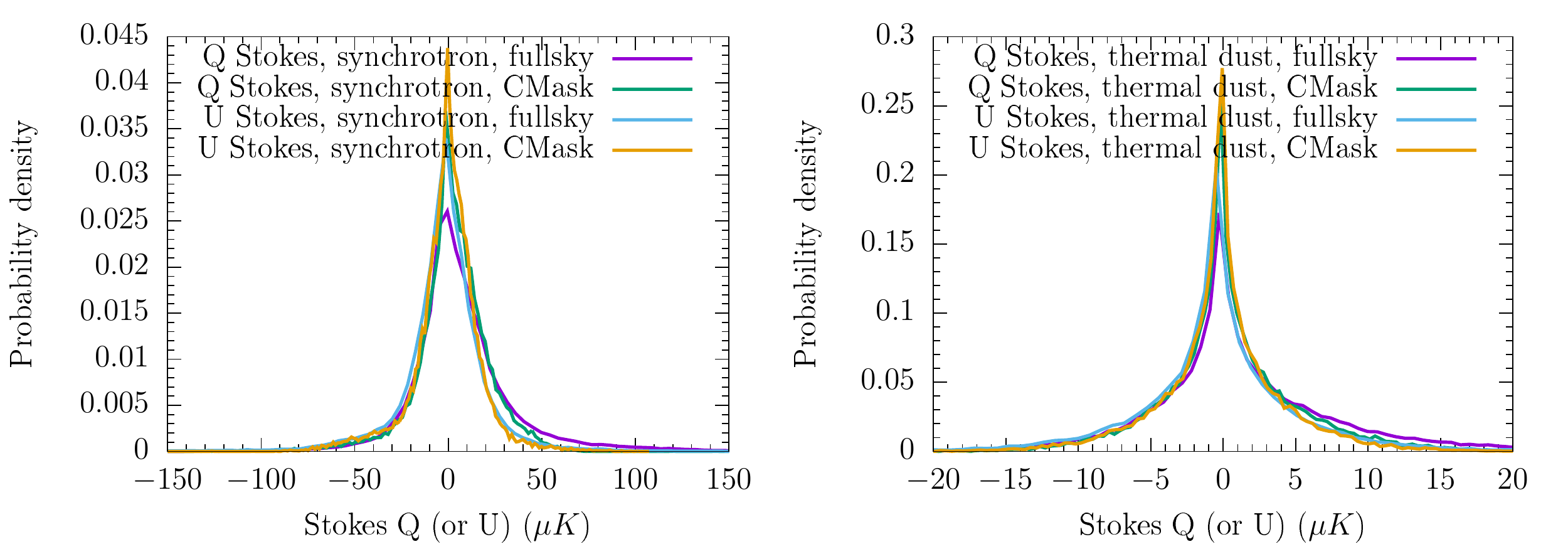}
\caption{Left panel: histogram of kurtosis values estimated from full sky as well as the region 
of the sky that survives after application of {\tt CMask} for synchrotron component for both 
Stokes Q and U polarization. Right panel: same as left panel but for thermal dust component. 
From these distributions one can easily infer about the non-Gaussian nature of these components 
for both Q and U polarization. }
\label{fg_hist}
\end{figure*} 
One of  basic principles on which the current work is based  is that the polarized foregrounds 
possess non-Gaussian properties. In this section we verify this assumption. To establish the 
non-Gaussian nature of polarized foregrounds we rely upon the polarized emission maps produced 
by the WMAP science team.  Both synchrotron and thermal dust emissions are polarized.  We verify 
non Gaussian nature of these components  using WMAP nine year  base-polarization model of 
synchrotron (at K band) and thermal dust (at W band) \citep{Gold2009, Gold2011}. Using these 
polarized foreground maps of these two frequencies we estimate the probability distribution of 
Stokes Q and U polarizations  of these two components from the entire sky region as well as 
from the region that survive after application of {\tt CMask}. We show these distributions  in 
Fig. \ref{fg_hist} which clearly show deviations from Gaussian nature. Although these  distributions 
appear symmetric they possesses sharp peaks which suggests non-Gaussian. The decay of 
probability  distributions on both  sides of the peak is concave upward for each case, which is 
in sharp contrast with respect to Gaussian distribution. Moreover, the distributions have long 
tails on both sides of the peak suggesting non-Gaussian properties of the underlying Stokes Q
and U samples. The kurtosis values for each of the cases plotted in Fig. \ref{fg_hist} are shown 
in Table \ref{fg_kurt}. From this table we note that synchrotron emission at K1 band shows stronger 
non-Gaussian properties than the non-Gaussian properties of thermal dust at W band for both 
polarizations and for both sky regions.  Moreover, for these regions the synchrotron Q Stokes 
signal shows stronger non-Gaussian than the corresponding U Stokes signal.  For the thermal 
dust component at W band, Q Stokes shows stronger non-Gaussian than U Stokes over the full sky. 
However, over the {\tt CMask} sky region, U Stokes signal of thermal dust is more non-Gaussian 
than the corresponding Q Stokes signal.

\begin{table}
 \caption{Kurtosis values of Stokes Q and U polarizations for foregrounds}
 \label{fg_kurt}
 \begin{tabular}{lllll}
  \hline
Sky region    &   Synchrotron  &     &  Thermal Dust &   \\  
  & Q Stokes  & U Stokes & Q Stokes & U Stokes \\  
  \hline 
Full sky & 149.193     & 54.009      &  11.302       & 8.132    \\  

{\tt CMask}   &  7.892     &  4.766      &   2.852     &    3.985  \\  
  \hline
 \end{tabular}
\end{table}

\section{Calibrating the non-Gaussian measure}
\label{Callib}
\begin{figure*}
\includegraphics[scale=0.65]{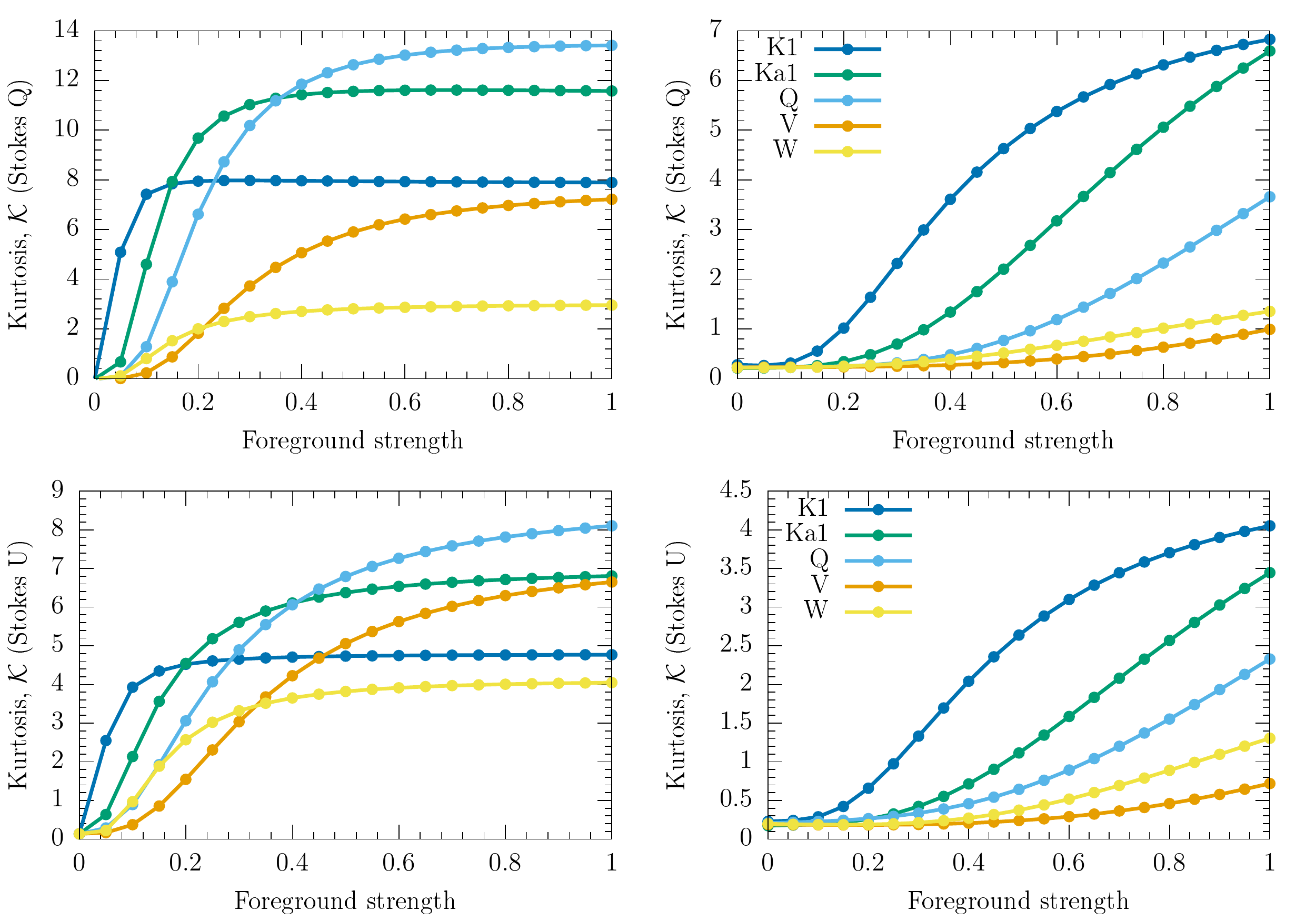}
\caption{Left  panel: variation of kurtosis with foreground strength added to randomly generated 
CMB Stokes Q and U polarization signals  for all five  WMAP frequency bands. Right panel: same as the 
left panel but in this case, detector noise consistent with  different WMAP frequency maps and 
varying levels of foregrounds  are added to the same CMB realization as in the left panel. All plots 
of this figure are obtained from {\tt CMask} sky region.}
\label{Kvsfg}
\end{figure*}

Given that in Section \ref{ng-pol_fg} we have empirically established the non-Gaussian nature of 
polarized foregrounds, in the current Section we attempt to understand how the kurtosis of 
diffuse foreground contaminated CMB polarization maps changes with the level of foreground 
contamination. We show this in Fig. \ref{Kvsfg} by plotting the values of kurtosis for foreground 
contaminated Stokes Q and U CMB maps as functions of foreground strength, separately for each of 
Stokes parameter maps, for two cases --  with and without detector noise contamination. To obtain 
CMB plus foreground  polarization maps at each of the WMAP frequencies with different levels of 
contaminations of foregrounds, we first form realistic foreground polarization maps at each of 
the five WMAP frequencies using the WMAP nine year base-model for synchrotron and thermal dust 
polarization components at $N_{\textrm side} = 512$ and at $1^{\circ}$ Gaussian beam resolution. 
For details of the polarized foreground model used in this work we refer to Section \ref{valid:model}. 
We randomly generate a realization of Stokes Q and U signal for pure CMB using Planck 2015 
best-fit power spectrum \citep{PlanckCosmoParam2016} at $N{\textrm side} = 512$ and also at 
$1^{\circ}$ Gaussian beam resolution. We add with the pure CMB signal foreground Stokes signals 
obtained above at various strength to form a number of CMB plus foreground only  polarization maps  
at  each of the five WMAP frequencies. We apply {\tt CMask} on each of these maps.  For each 
frequency  band we find the sample kurtosis values from the surviving pixels as function of the 
increasing foreground strength. The variation of kurtosis values with the increase in foreground 
contamination is shown in the left panel of Fig. \ref{Kvsfg} for CMB plus foreground only case.  
Foreground strength, as plotted in the horizontal axes of this figure simply represents the 
multiplicative factors that were applied to the net foreground signal at each frequency before adding 
them with the CMB. 
As we see from the left  panel of this figure the kurtosis values increases quickly as the foreground 
strength increases from zero for both Stokes Q and U signal. For larger foreground strength the 
kurtosis values for all frequency maps almost saturates (with minor fluctuations not visible in the 
scale of these figures). Such behavior may be explained by noting that by increasing foreground 
strength in these figures  CMB polarization signal becomes increasingly weaker compared to the 
foregrounds. The saturation property of the graphs then follows from a simple property of  the 
kurtosis of a set of samples that kurtosis remains invariant by simply scaling all sample values by a 
unique constant. An important observation from this figure is the  presence of a global minimum in the kurtosis 
value for all WMAP frequency maps when the foreground contamination becomes absent. The presence of the 
global minimum also seen in the cases of foreground and detector noise contaminated CMB polarization 
maps, as shown in the right panel of Fig. \ref{Kvsfg} justifies globally minimizing  kurtosis to 
eliminate the non-Gaussian foreground contaminations. The right panel of Fig. \ref{Kvsfg} shows the 
variation of kurtosis for foreground and detector noise contaminated polarized CMB maps after application 
of  {\tt CMask}. For details of forming the the noise realizations we refer to Section \ref{valid:model}. 
For each frequency map only the foreground strength is varied with the CMB and detector noise being 
compatible to LCDM signal and WMAP frequency maps respectively and corresponding kurtosis values are 
plotted as functions of foreground strength. In interesting difference of this plot with the corresponding 
ones of the left panel is that  unlike the top panel the kurtosis values increases even with the largest 
level of foreground contamination for all frequency maps and for both Stokes Q and U polarization. This 
can be explained by noting that although polarized CMB signal is much weaker  
than the polarized foreground contamination  detector noise contamination in WMAP frequencies is not negligible 
compared to the foreground contaminations. The rise of kurtosis values with increasing the foreground strength 
becomes slower for the detector noise contaminated case compared to the case when  detector noise is absent . 
This can be explained by noting that both the detector noise and CMB are Gaussian, and noise is stronger than CMB 
but comparable with the foregrounds. Another important observations from this figure for each frequency maps 
and for both Stokes polarizations overall the values of kurtosis becomes smaller when the maps have detector noise 
contamination added in them. This is expected since in presence of detector noise foreground removal becomes less 
effective than the noise free case.  Finally we note that it is worth listing the kurtosis values from the {\tt CMask} 
sky region without any foreground addition but for both pure CMB and  pure CMB  plus  detector noise addition
for both Stokes parameters for the particular CMB and detector noise realizations used to make the plots 
Fig. \ref{Kvsfg} . The kurtosis value 
of the particular CMB realization for Stokes Q and U maps are respectively $-0.0224$ and  $0.1349$ . Adding detector 
noise to Q maps (with zero foreground level) increases kurtosis values that depends upon K1, Ka1, Q, V and W frequency bands 
as follows,  $0.2786, 0.2113,  0.2303, 0.2261, 0.2168$. Similarly,
for the U polarization (with zero foreground level) the kurtosis values increase with addition of detector 
noise to pure CMB U signal ($0.2298, 0.1742, 0.2086, 0.1889, 0.1971$ respectively for K1 to W bands.)
Such increase in the value of  kurtosis with addition of detector noise may be explained by noting that 
kurtosis values show larger variance when  detector noise which is a strong signal 
than the pure CMB polarization signal to the pure CMB Stokes Q and U polarization signals.

\section{Non-Gaussian Nature of WMAP maps}
 \label{inpmap_kurt}

\begin{figure*}
\includegraphics[scale=0.8]{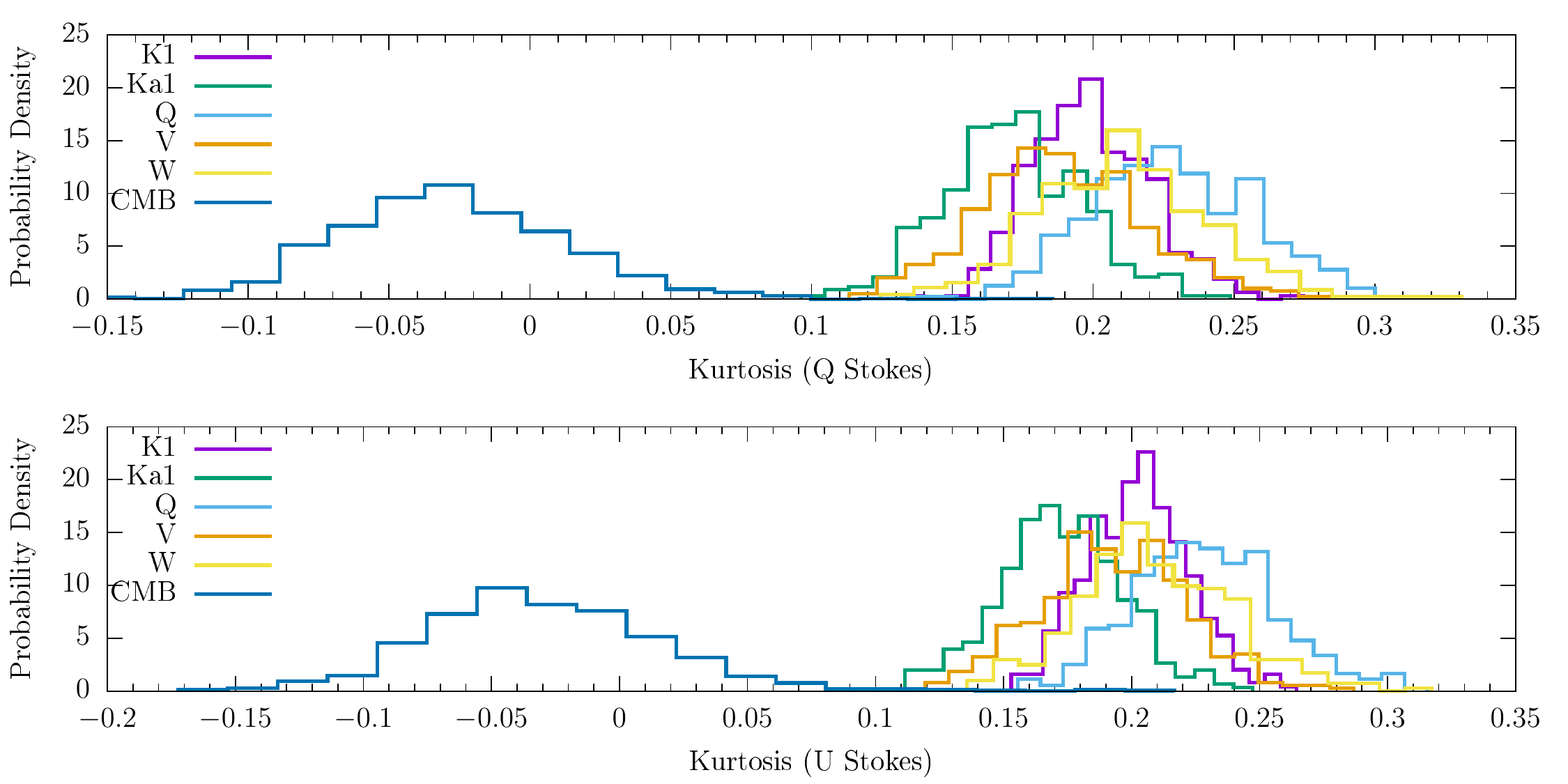}
\caption{Top panel: histograms of kurtosis values from region of the sky that survives after application of 
{\tt CMask} for pure CMB and pure CMB plus detector noise (compatible to WMAP frequencies) maps for Stokes Q 
parameter. Bottom panel: same as  top panel, but for U Stokes parameter.  Clearly, addition of detector 
noise causes the kurtosis values to deviate towards the positive sides
for both the Stokes parameters.}
\label{hist}
\end{figure*}  

\begin{table}
 \caption{Kurtosis values of input Q and U Stokes maps estimated from full sky as well as from the {\tt CMask} sky-region.}
 \label{inp_kurt}
 \begin{tabular}{llllll}
  \hline
 Stokes and    &  K1  & Ka1 &  Q & V &  W  \\
 sky-region   &      &     &    &   &      \\
  \hline
 Q, Full sky        & 677.946 &  1178.631 & 1279.929 & 212.516 & 3.283 \\

 Q,  {\tt CMask}       &  17.751  & 8.154 & 3.444   &   0.251   & 0.418 \\
  \hline
U, Full sky        & 69.836 &  36.613 & 13.533 & 0.898 &  0.29 \\

 U,  {\tt CMask}       &  5.764  & 1.806 & 0.902   &   0.311   & 0.293 \\ 
  \hline
 \end{tabular}
\end{table}

Following our discussions about strong non-Gaussian nature of polarized foregrounds 
from the view point of density analysis in Section \ref{ng-pol_fg} and after providing 
justifications of using kurtosis statistic as a measure of non-Gaussian property 
of polarized foregrounds (plus detector noise) contaminated CMB map in Section \ref{Callib} 
in this Section we turn our attention to  quantify  non-Gaussian properties are inherent 
into the five input WMAP maps using the kurtosis statistic.  We estimate sample kurtosis 
of both Q  and U polarizations for all five input maps  at different WMAP frequencies for 
the  full sky as well as from the region of the sky that survive after applying {\tt CMask}.  
We show these kurtosis values in  Table \ref{inp_kurt}. Although, K1 band has the maximum
power for both EE and BB power spectrum \footnote{This may be easily seen by plotting the full sky 
EE and BB power spectra from the five input maps discussed in Section \ref{inputs}.}, we note that the Q band has the maximum non-Gaussianity. This 
clearly indicates that the non-Gaussian measure of this work identifies some contamination 
which are not detected in simple power-spectrum based measure of contamination. The W band 
gets minimum kurtosis amongst five bands for the full sky case, since it has lowest synchrotron 
emission level and a thermal dust emission level that is also comparable  to detector noise 
level of W band. After application of {\tt CMask} V band gets minimum kurtosis (unlike W band
for the full sky case)  which is consistent with the expectation that V band is minimally 
foreground contaminated within the observation window of WMAP. Moreover, as seen from the 
second row of this table  the kurtosis of Q polarization estimated from {\tt CMask} sky region 
gradually decreases from K band to V band. This particularly implies the excess contamination 
for Q band that is observed in the measured kurtosis of the full sky  Q band map, 
arise from the sky region that is outside the {\tt CMask}. For the U polarization the kurtosis 
values decreases gradually from K1 band to W band for both full sky and {\tt CMask} sky. From the 
systematically larger kurtosis values of Q over  U polarization for the full sky we conclude 
that the foregrounds in U polarization possess less non-Gaussian properties  than the foregrounds in Q 
polarization. Moreover, for V and W bands kurtosis values for U polarization estimated
from {\tt CMask} region, are comparable. This likely implies that detector noise for V and W band  
for U polarization are at least as strong as the thermal dust emission which is the dominant 
foreground of these frequencies.

\section{Kurtosis variation without foreground}
\label{kurt_no_fg}
In section \ref{Callib} we have discussed  variation of kurtosis with increase in foreground strength for
pure CMB plus foreground contaminated maps and pure CMB plus both detector noise and foreground contaminated
maps for each of the five WMAP frequency maps from the region of the sky that survive after applying {\tt CMask}. 
In practice, to correctly interpret value of measured kurtosis for both foreground and detector noise 
contaminated case we also need to know typical values and variation of kurtosis for pure CMB  plus detector 
noise case in the absence of any foreground contamination. For this purpose we simulate $1000$ realizations 
of pure CMB Stokes Q and U maps at $1^\circ$ beam resolution and at $N_{\textrm side} = 512$. We apply {\tt CMask}
on each of these realization and estimate kurtosis values for both Q and U polarizations from the  
surviving pixels. Histograms of  these kurtosis values are  shown in the top and bottom panel of Fig. \ref{hist}. 
The sample mean ($\hat \mu$)  and standard deviation ($\hat \sigma$)  for the Q Stokes for pure CMB maps from the 
{\tt CMask} sky-region  are respectively 
$ -0.01251$  and    $ 0.04268 $. The corresponding values for U polarizations are $-0.01257$ and   
 $0.04818 $ respectively. As expected, the sample mean (or standard deviation) values are nearly 
identical for the case of Q and U polarization, which is expected since both Q and U Stokes parameters
are equally preferred for CMB anisotropies.  
The histograms of kurtosis values estimated from {\tt CMask} region of Q and U polarization maps for CMB plus  
detector noise case are also shown in Fig. \ref{hist} for all the five WMAP frequencies \footnote{These histograms were obtained from 400 simulations 
of CMB plus detector noise maps corresponding to  each of the five WMAP frequencies. We refer to Section \ref{valid:model}
 for a discussions about generating the noise simulations.}. The mean sample 
kurtosis for Q Stokes parameter are $0.2087, 0.18, 0.2378, 0.202$ and  $0.2118$ respectively 
for K1, Ka1, Q, V and W band maps. The corresponding sample mean values for U polarization are respectively 
$0.2071, 0.1801, 0.2386, 0.1986$ and  $0.2123$. The standard deviation for both Q and U polarizations 
are approximately independent on any of the two  polarization types and frequency maps and varies approximately 
between $0.02$ and $0.03$.   Comparing the sample mean values of kurtosis for pure CMB  and CMB plus detector 
noise cases, we conclude  addition of detector noise makes the kurtosis to 
significantly deviate towards the positive values.

\section{Methodology}
\label{method}
A well known  property of the usual ILC method (either in pixel space or harmonic space at each multipole $\ell$) 
that relies upon estimating the weight factors by minimizing the empirical variance of cleaned map is that weights can be directly 
computed using an analytical expression. The analytical estimation of weights is possible in that case since
the empirical variance of the cleaned map is a quadratic function of the weight factors. As a result, the best-fit 
weights, which are solution obtained by differentiation of the variance and suitably setting the result to zero 
by employing a Lagrange's multiplier approach, satisfy an equation which is linear in the weights themselves 
(e.g., see Eqn A3 in appendix A of \cite{Saha2008}). Because of this linear equation weights can easily be 
obtained in terms of an analytical equation (e.g., \cite{Tegmark96, Tegmark2003, Saha2006,Saha2008,Saha2017})
in ILC methods that minimize variance. In the current work we however minimize kurtosis as a measure of non-Gaussian 
properties inherent in the data. Since the kurtosis of the cleaned ILC map is highly non-linear function of 
weights, e.g., see Eqn. \ref{kurt_cmap},  (unlike the {\it minimally non-linear} quadratic function as in the case 
of variance minimization) differentiating this kurtosis with respect to the weights following the Lagrange's undermined 
multiplier's approach does not result in  a linear equation of weights. Hence an analytical expression for weights 
for our current work does not seem feasible. Instead as in \cite{Saha2011} we develop a code (hereafter called 
{\tt PolG} to solve for weights that minimizes Eqn. \ref{kurt_cmap} 
by employing a non-linear minimization algorithm due to Powell as in \cite{nrc}.

\section{Results}
\label{results}
\subsection{Foreground cleaned Stokes Q and U maps}

\begin{figure}
\includegraphics[scale=0.68]{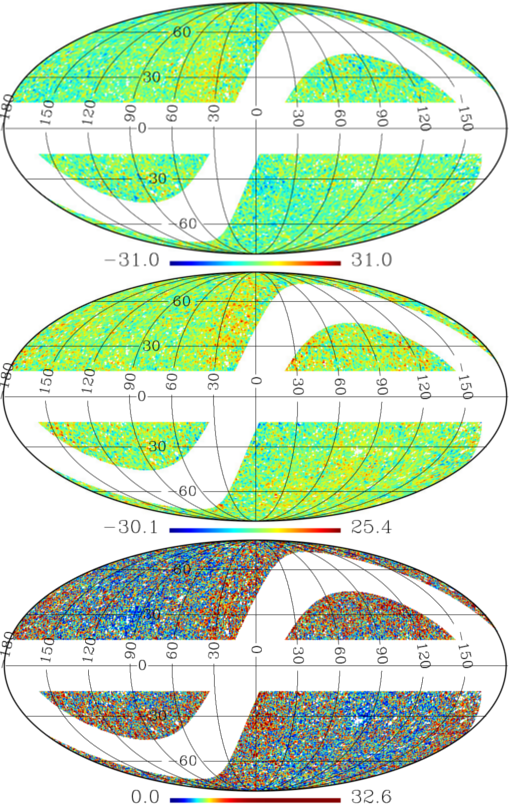}
\caption{Top: foreground cleaned Q Stokes map obtained by our method. Middle: foreground cleaned 
U Stokes map. Bottom: map of $P = \{Q^2 + U^2 \}^{0.5}$, using Q and U maps of top and middle panels.
The cleaned Q and U Stokes maps of this work together are referred to as {\tt QUGMap}.}
\label{CMap}
\end{figure}

\begin{figure*}
\includegraphics[scale=0.66]{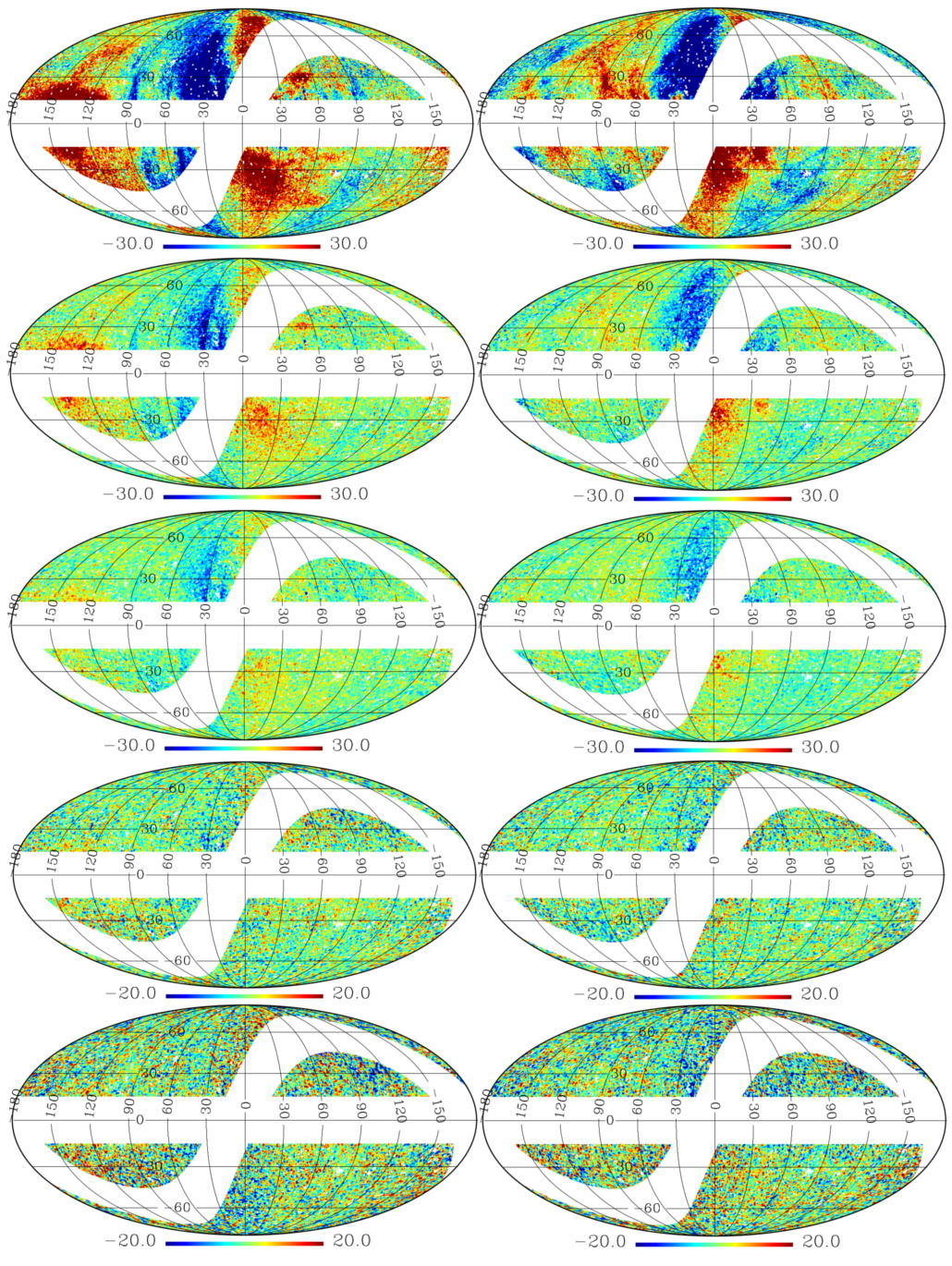}
\caption{Left column: from top to bottom, input Stokes Q maps for WMAP K1, Ka1, Q, V and W band 
respectively (after applying {\tt CMask}). Right column: same as left column, but for U Stokes parameter. }
\label{InpMasked}
\end{figure*}

First, we apply {\tt CMask} obtained in Section \ref{mask} to Stokes Q polarization maps of the five 
WMAP frequency bands which have a common resolution of $1^\circ$, as discussed in Section \ref{inputs}.  
Using the {\tt PolG} code we then minimize kurtosis  over  the surviving 
pixels of five Stokes Q polarization maps. Since WMAP V  band is expected to have minimum 
foreground contamination from the {\tt CMask} sky region, (e.g., see table \ref{inp_kurt}), 
we choose V band weight to be unity and all other  frequency band weights to  be zero 
as the initial values of the weight factors.
For the Q polarization  weight vector ${\bf W}$ becomes 
$\left(-0.237422,  0.376352,  0.136837,  0.691807,  0.032427\right)$
for the minimum value of kurtosis $0.188642$ 
This is clearly less than the kurtosis values of all the 
five input WMAP Q polarization maps, estimated after application of {\tt CMask} (e.g., see Table \ref{inp_kurt}). The {\tt PolG}
code generates, the maximum weight for V band, since it is expected to be the minimally foreground 
contaminated frequency band over the {\tt CMask} region as discussed in Section \ref{inpmap_kurt}. 
It is interesting to  note that both  Ka and Q bands get positive weights, whereas, at the low frequency side, 
K band gets a negative weight. This is because strong synchrotron signal at the WMAP 
low frequency side (e.g., at 23, 33 and 41 GHz) is canceled by assigning negative weight 
to K1 band, where the synchrotron contamination is strongest and positive weights to Ka and Q bands 
where synchrotron contamination becomes gradually weak. The highest frequency WMAP 
frequency band, on the other hand gets, a very small weight since this map is dominated by the 
thermal dust emission. From table \ref{inp_kurt} we note that for U polarization both V and W band 
have similar kurtosis $\sim 0.30$ from {\tt CMask} sky. This kurtosis value is well within the range of kurtosis values 
we obtain from the simulations of pure CMB and detector noise (e.g., see Fig. \ref{hist}). 
We conclude from this both W band V band U polarization maps which likely contains detector 
noise contamination that is at the least comparable or in practice stronger compared to the 
foreground emission levels of {\tt CMask} regions of these frequency bands. Because of larger detector
noise contamination the V and W  frequency bands provide little leverage in removing foregrounds. 
Also, in practice excluding these two frequency bands from the   analysis is not an option 
since one would like use a wide frequency coverage for better estimation of CMB 
signal removing foregrounds, particularly when data becomes noise dominated. We therefore
do not use U polarization maps to estimate the weights. Instead, we simply use the weights 
obtained for Q Stokes map to obtain a foreground cleaned U polarization map from the {\tt CMask}
region.  The kurtosis value of cleaned U polarization map obtained in this way is 
comparable  ($\sim 0.333$) to kurtosis of input V and W bands (from {\tt CMask} region). We 
show the cleaned Q and U Stokes maps in the top and middle panel of Fig. \ref{CMap}. The 
corresponding map of magnitude, $P = \{Q^2 + U^2\}^{0.5}$  of polarization vector, is 
shown in the bottom panel of this figure.  Clearly the polarization map shows presence of 
residual detector noise contamination around the region along the ecliptic plane. Presence of 
some residual foregrounds are visible near north galactic spur and on the southern galactic plane. Our
cleaned Q and U Stokes maps together hereafter will be referred to as {\tt QUGMap}.

\subsection{Comparison with Input Stokes Q and U  Maps}
A direct way to demonstrate the efficiency of foreground minimization may be obtained 
by comparing the input foreground (and detector noise contaminated) maps with the 
final foreground cleaned maps by our method for both the Stokes parameters. 
We show all five input WMAP Q and U Stokes polarization maps after application of 
{\tt CMask} in Fig. \ref{InpMasked}. From top to bottom the left column of this  figure 
shows Q Stokes maps from K1, KA1, Q, V and W band respectively. The right panel shows 
the same, but for U polarization. Comparing this figure with top and middle 
panel of Fig. \ref{CMap} we see that our method effectively minimizes
strong foreground that is  present  in K1 or Ka1 band. Foreground contamination clearly 
visible from north polar spur region and immediately below the galactic plane 
 for WMAP Q band for both the Stokes maps. However, 
such features is absent in our foreground cleaned map which implies efficient foreground 
minimization by  {\tt PolG}. For  W band the input maps show features that 
are more consistent with contamination due to both detector noise and foregrounds (thermal dust)
and a visual inspection  of foreground minimization  algorithm is not reliable.  The 
efficiency of foreground removal with respect to these input maps may be clearly demonstrated 
by noting the standard deviation and kurtosis values of all the maps.        

\subsection{Difference with Input Polarization Maps}
\begin{figure}
\includegraphics[scale=0.66]{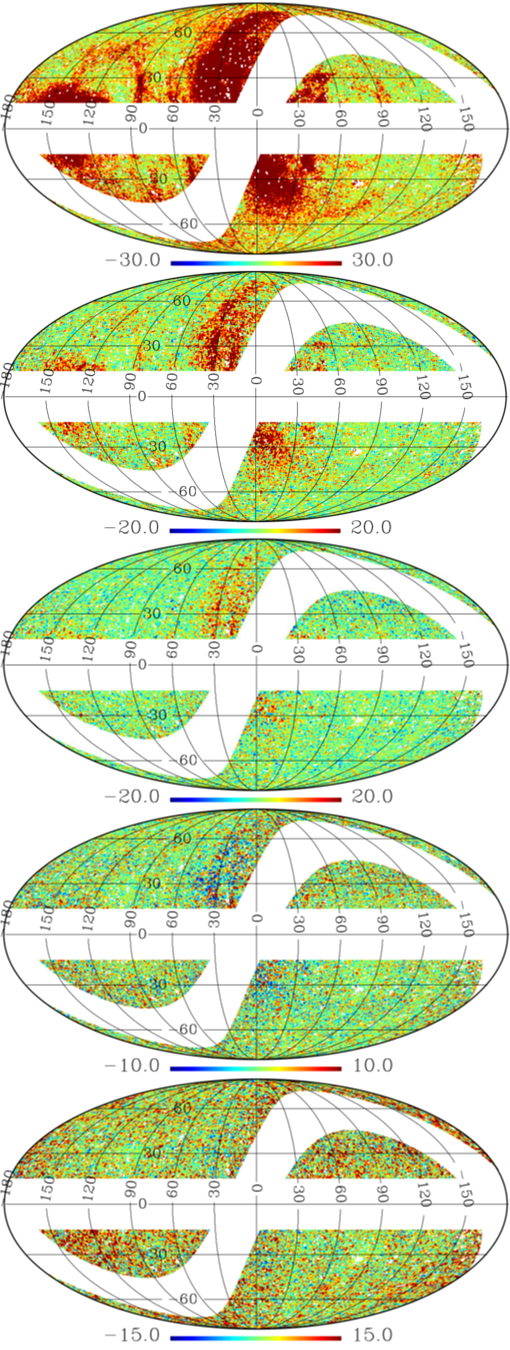}
\caption{Difference of magnitude of polarization vectors  of the input maps and output foreground 
minimized map obtained  by our method. From top to bottom images  represent difference maps for K1 to 
W band in the increasing order of the frequency.}
\label{PMaps}
\end{figure}

What is the difference between the magnitude of polarization vectors estimated from input 
maps and our Stokes Q and U cleaned maps? Such difference maps are of particular interest since 
they show the performance of the algorithm in reducing foreground power from the 
regions of the sky where foreground are strong compared to both detector noise and  the 
expected level weak CMB polarization signal. We show these difference maps in Fig. \ref{PMaps} 
from top to bottom for K1 band to W band in the increasing order of their 
frequencies. From the topmost image of  this figure we see that  strong
foreground contamination of K1 band is minimized in both the northern and southern 
galactic hemisphere. Particularly visible from this figure the efficiency of 
our algorithm from the left hand side of galactic plane (on both hemispheres) and 
the ring like structure in the northern galactic plane. The  difference maps for 
each of Ka1 and Q bands shows efficient foreground minimization near the north 
galactic spur region. For V and W bands the difference maps tend to loose visual 
presence of foreground patterns since foregrounds becomes comparable to the 
detector noise level.   

\subsection{Variance Analysis}
\begin{figure*}
\includegraphics[scale=0.7]{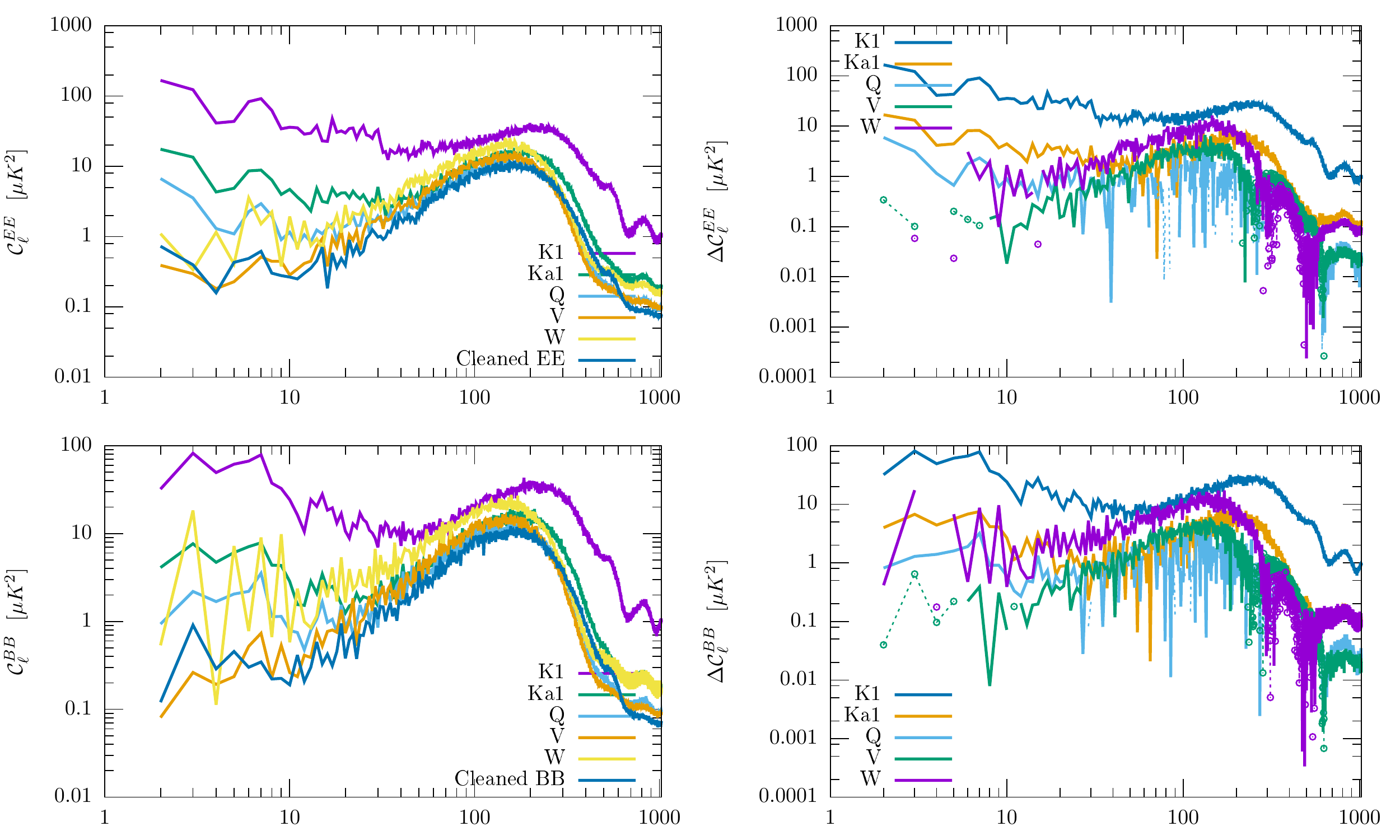}
\caption{Top left and bottom left: figures showing the  EE and BB power spectra respectively obtained 
from {\tt CMask} region from cleaned Stokes Q and U maps of this work along with the corresponding spectra
from five input maps (as discussed in Section \ref{inputs}) after application of {\tt CMask}.  The vertical 
axes of both the figures show the `reduced spectrum', $\ell(\ell+1) C_{\ell}/2\pi$. Top right and bottom right: 
The absolute difference of reduced EE and BB spectra respectively, estimated from our \{Q, U\} cleaned maps 
and five input maps from {\tt CMask} region.  See Section \ref{pow_spec_inp} for details. }
\label{inp_out_cls_data}
\end{figure*}

It is interesting to note that, the standard deviation values of cleaned Q ($5.50\  \mu K$) 
and U ($5.48\  \mu K$)  polarization maps from {\tt CMask} region  match very closely to each 
other. The standard deviations of five input Q polarization maps from the same sky region 
 are $21.14, 8.94, 6.48, 6.25$ and $ 8.44\  \mu K$ from the K1 band to W band, respectively. 
The corresponding values for U polarizations are respectively 
$19.16, 8.33, 6.35, 6.25$ and $8.35\  \mu K$. This shows that our foreground removal method reduces 
the sample standard deviation of the uncleaned maps by at least $0.75\  \mu K$ for Q Stokes 
polarization and  $0.77\  \mu K$ for the U polarization, when we compare with the corresponding 
standard deviation of the uncleaned V band maps. When compared with the K1, Ka1 and W band the  
reduction in strength of contaminations are even stronger, indicating the efficient 
performance of the method. In this context we note that since CMB  
polarization is a weak signal than the temperature counterpart,  pure CMB  
(both Q or U Stokes maps, after smoothing by $1^\circ$ Gaussian  window function)  has a mean standard 
deviation value of only $\sim 0.55\  \mu K$ over the same {\tt CMask} sky region with a sample standard deviation 
$0.018\  \mu K$. Even, considering the performance 
of foreground removal as compared with the two lowest variance uncleaned V or W bands,  one may therefore 
conclude that  a reduction of standard deviation by $\sim 0.75\ \mu K$ using {\tt PolG} code  amounts 
to significant reduction of contamination compared to the weak CMB polarization signals.   It is interesting to note that for both Q and  
U polarization the standard deviations of cleaned Q and U maps are less than the minimum 
standard deviations of all five input maps, although our foreground minimization algorithm 
is independent on variance minimization.  

\begin{figure*}
\includegraphics[scale=0.72]{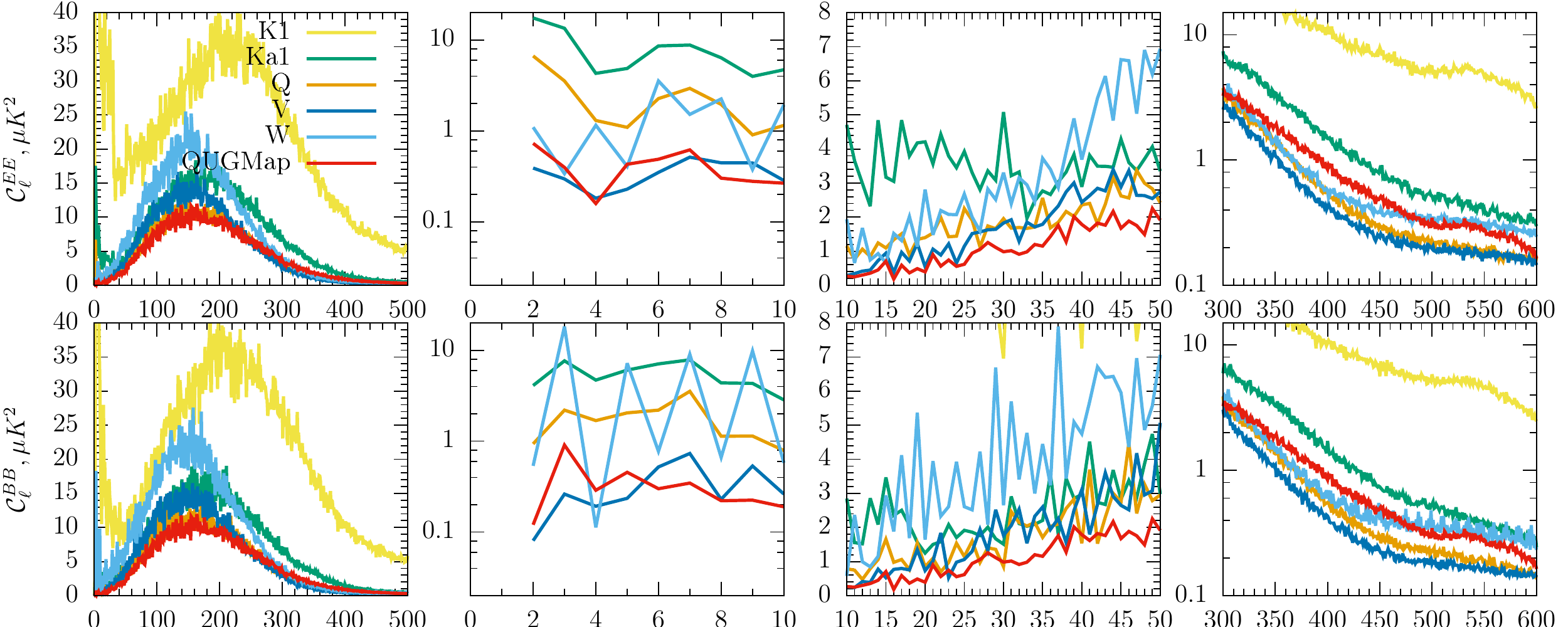}
\caption{Comparison of EE (top panel) and BB (bottom panel)  power spectra,  
$\mathcal C_{\ell} = \ell(\ell+1) C_{\ell}/2\pi$, obtained from {\tt QUGMap}
with the corresponding EE and BB spectra of all input maps  used in this work at different WMAP frequencies, for  different ranges of multipoles. 
The K1 band spectra remain outside the vertical plot range shown in  figures (1,2), (1,3) and (2,2). 
All spectra are estimated from {\tt CMask} region.}
\label{CompCls1}
\end{figure*}

\subsection{Power Spectrum}
\subsubsection{Comparison with Input Power Spectra}
\label{pow_spec_inp}
An effective way to discuss the efficiency of foreground minimization by a foreground 
removal method is to study the spectra obtained from the final cleaned maps produced 
by the method and compare them with the corresponding spectra estimated from the input 
uncleaned maps. We compare  EE and BB power spectra of {\tt QUGMap} obtained from {\tt CMask} 
region with the corresponding spectra of the foreground 
contaminated Stokes Q and U maps corresponding to five WMAP frequencies (see Section 
\ref{inputs} for a discussion about how these maps are generated). We show the 
results in Fig. \ref{inp_out_cls_data}. The top left plot of this figure shows our 
cleaned EE power spectrum, $\mathcal C^{EE}_{\ell} = C^{EE}_{\ell}\ell (\ell+1)/2\pi$ 
along with the corresponding EE spectra estimated from the five input frequency maps,
all spectra being estimated from the {\tt CMask} sky region.  The top left figure shows the
overall variation of the EE spectra for the entire range of multipoles. The top right plot of Fig. \ref{inp_out_cls_data} 
shows absolute magnitude  of difference of input and cleaned EE spectra.   
For any of the five difference spectra of top right figure we use  solid lines if the value of cleaned EE spectrum 
is less than the corresponding input  EE spectrum of the corresponding  WMAP input 
map at any given multipole. In such a case, some foregrounds have been removed by 
our method from the polarization map of the concerned input Stokes Q and U maps,    
at the particular multipole. If, on the other hand, the cleaned EE spectrum is more than 
corresponding EE spectrum of any input map at a multipole the corresponding values of 
the  absolute difference is plotted in points or dashed line. Such points represent  multipoles 
where our foreground removal has not become very effective. As seen from the top right plot 
our method reduces a significant amount of power  from input K1 and Ka1 frequency bands for  
all multipoles. The cleaned EE spectrum is less than Q band EE spectrum for $\ell \lesssim 40$.
For V and W bands, except for lower few multipoles our method effectively removes 
foreground power up to intermediate multipoles.  The bottom left figure of Fig. \ref{inp_out_cls_data} 
represent  BB spectra ($\mathcal C^{BB}_{\ell} = C^{BB}_{\ell}\ell (\ell+1)/2\pi$) of {\tt QUGMap} and 
all five input maps from {\tt CMask} region. The bottom right figure shows the absolute difference 
of BB spectrum of our cleaned map and any of the input WMAP polarization maps. Like the top right 
figure, the solid lines of the bottom right figure shows the multipoles and frequency bands for 
which our cleaned map has lower foregrounds. The points  or the dashed lines represent the 
multipoles for which our foreground removal is not very effective.  In top panel of Fig. \ref{CompCls1}
we compare the cleaned EE spectra  with the uncleaned EE spectra of five input frequency bands 
from {\tt CMask} region, for different ranges of multipoles. Bottom panel represents corresponding 
plots for BB spectra. The color codes of the sub-figures of Fig. \ref{CompCls1} are identical to that 
of top left figure. The (1,1) (following (row, column) convention)  and (2,1) figures  shows the overall 
variation of EE and BB spectra up to multipole $\ell = 500$. (1,2) and (2,2) figures show that our cleaned 
EE and BB spectra have less power than K1, Ka1, Q frequency band input polarization maps for $\ell \le 10$. 
Both EE and BB spectra are less than corresponding W  band spectra at almost all multipoles $\le 10$.
The V band EE and BB spectra are comparable, or less than   the cleaned EE and BB spectra at the 
lowest few multipoles,  however the cleaned spectra have less power than all WMAP input maps soon 
as $\ell$ increases beyond multipole 10. This is shown in figures (1,3) and (2,3). At higher multipoles
$\ell \ge 220$ the cleaned spectra again possesses more power than V band spectra (e.g., (1,1)  and
(2,1) figures). At even higher multipoles the cleaned spectra have more power than Q, V and W band 
input maps, but less power than K1 and Ka1 band input maps, as shown in figures (1,4) and (2,4).             

We note in passing that the BB power spectra obtained from power spectra are merely
an indication of residual foregrounds and detector noise contamination. The measured
BB spectra  should not be interpreted to be caused by the primordial  CMB signal
since primordial BB signal is expected to be of much lower magnitude, much lower than
even CMB EE power spectrum.

An important phenomena in the context of EE and BB power spectrum estimation
from partial sky is leakage of EE signal to BB and vice versa \citep{Tegmark2001, Lewis2002, Gabor2003}.
The leakage requires careful analysis to extract accurate primordial CMB signal from the mixture
of EE and BB spectra obtained from the partial sky. In particular \cite{Samal2008}
suggest that the leakage can be avoided if one converts the full sky Stokes Q and U polarization maps
to E and B mode maps at the beginning of the analysis. In the current work, however, we
have chosen to apply {\tt CMask} on the input WMAP  Q and U Stokes maps at the beginning of the
analysis, without converting the full sky  Stokes maps to E and B maps at any stage of our analysis.
This necessarily implies that there would be leakage of power between EE and (foreground plus
detector noise) BB mode in the all the spectra shown in Fig. \ref{inp_out_cls_data}. However, for
the mere comparison of residual power in different cleaned maps the leakage between EE and
BB modes is not a concern and therefore we avoid reconstruction of true EE (or BB)
that will be measured without leakage, from the partial sky estimates of EE and BB spectra
shown in Fig. \ref{inp_out_cls_data}.  The partial sky EE and
BB spectra of this figure merely represent the measure of residual contaminations and not any
accurate estimate of primordial CMB EE or BB power spectra.

\subsubsection{Comparison with WMAP foreground-reduced Power Spectra}
\label{pow_spec_wmap}
\begin{figure*}
\includegraphics[scale=0.72]{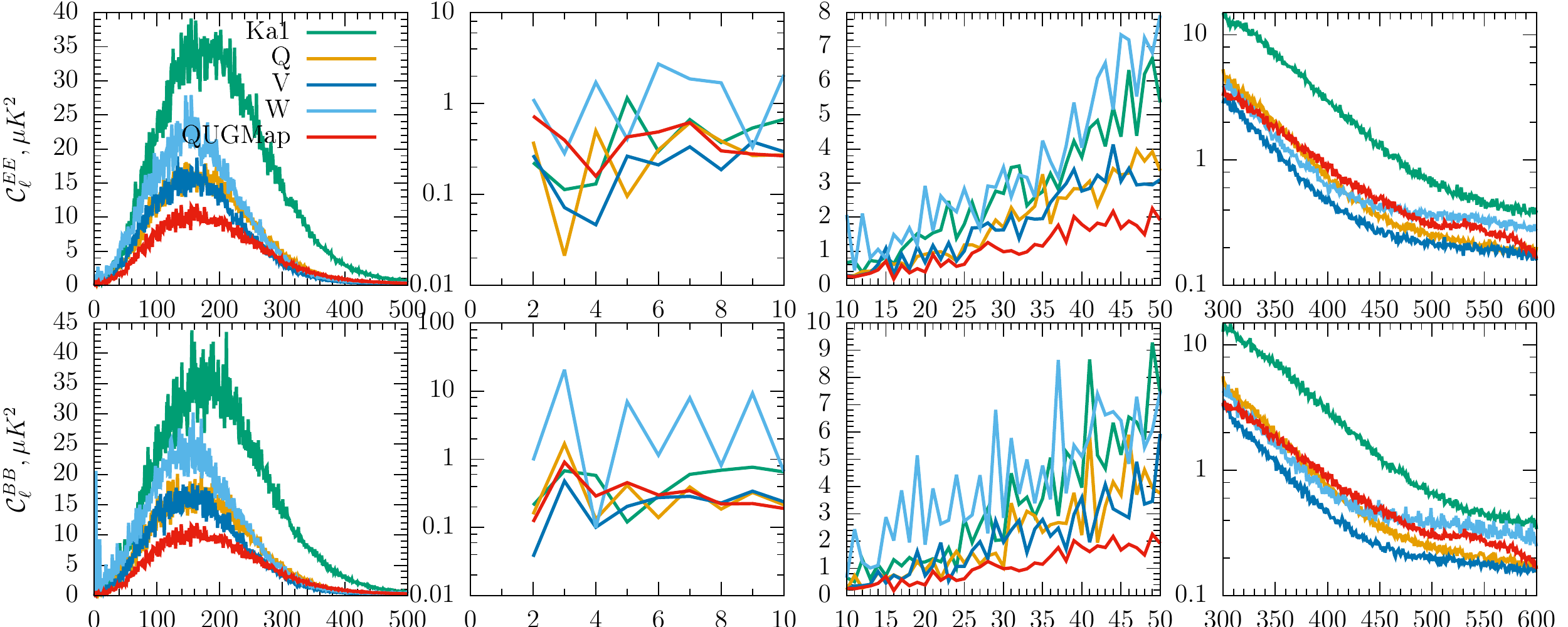}
\caption{ Comparison of EE (top panel)  and BB (bottom panel) power spectra,  
$\mathcal C_{\ell} = \ell(\ell+1) C_{\ell}/2\pi$, obtained from {\tt QUGMap} 
with the corresponding spectra estimated from all four WMAP foreground reduced  polarization maps  for  
different ranges of multipoles. All spectra of this plot have been estimated from {\tt CMask} 
sky region.
}
\label{CompCls}
\end{figure*}
WMAP Science team have produced a set of foreground cleaned CMB Stokes Q and U 
polarization maps for each DA except the DA for the K1 band \citep{Page2007, Bennett2013}.
As described in \cite{Page2007} for polarization foreground removal synchrotron
emission is modeled by using the K1 band Stokes Q and U maps taking care of weak CMB signal  
of this frequency band. The authors then, first, model thermal dust intensity, $I(\hat n)$,
at $94$ GHz using FDS model eight of \cite{Fink1999}.  Using this intensity model
\cite{Page2007} then obtain Stokes Q and U template maps using Eqn 15 of their paper.
The templates then fitted with the data from Ka1 to W band DA maps by minimizing 
the $\chi^2$ defined by Eqn. 19 of their paper.

We compare  EE and BB power spectra obtained from {\tt CMask} 
region of {\tt QUGMap}  with the corresponding spectra of the foreground 
cleaned Stokes Q and U maps provided by WMAP science team. For this purpose we 
first smooth the Ka1 band Q, U Stokes maps to $1^{\circ}$ Gaussian beam 
resolution at $N_{\textrm side} = 512$ performing the smoothing operation in the 
multipole space using ratio of the window function of $1^{\circ}$ Gaussian beam 
and the native beam function of the WMAP Ka1 band.  For each Q, V and W  bands 
multipole DA maps available. We convert each of these DA maps to $1^{\circ}$ Gaussian 
beam smoothed maps following procedure similar to used for Ka1 band. Moreover, for 
the each of Q, V and W bands we average the smoothed DA maps over the available 
DA maps at the corresponding band. This results in a total of four foreground cleaned 
maps for each of Q and U Stokes parameter for Ka1, Q, V and W band respectively.
We apply these maps by the {\tt CMask} and estimate their EE and BB power-spectra 
along with the power spectrum,  obtained from our {\tt QUGMap}, from the same sky region. We show these spectra 
at the top and bottom panel, respectively, of Fig. \ref{CompCls} for different 
multipole ranges. The vertical axes of each of the images of this plot represents 
$\mathcal C_{\ell} = \ell(\ell+1) C_{\ell}/2\pi$. Overall variation of different
spectra between the multipole range $\ell \le 500$ is shown in the (1,1) and (2,1) figures.
The (1,2) and (2,2)  figures  show all the spectra at the lowest multipoles, $2 \le \ell \le 10$. 
At $\ell =2 $  and $3$, EE power from {\tt QUGMap},
becomes comparable to WMAP W band cleaned map, but larger
than power in other cleaned WMAP  maps at these multipoles. We can expect that some residual 
contamination is present in our {\tt QUGMap} (and WMAP cleaned W band map) at this multipole. 
However, $\mathcal C^{EE}_{\ell}$ soon catches up  with the corresponding spectra of all 
other cleaned WMAP maps as $\ell$ increases. At $\ell =4$, $\mathcal C^{EE}_{\ell}$ from  
{\tt QUGMap} becomes lower than corresponding power of Q and W bands, comparable to Ka1 band EE 
power and larger than V band EE power. At $\ell = 10$ EE power from {\tt QUGMap} becomes comparable 
to corresponding power of cleanest V band cleaned map produced by WMAP. Similar pattern is also 
seen from the BB power spectra shown in figure (2,2). In the 
case of BB spectra power in our cleaned spectrum becomes less than cleanest WMAP V band BB 
spectrum  starting from multipole $\ell \sim 8$.  As shown in (1,3), (2,3), (1,1) and (2,1) figures  
of Fig. \ref{CompCls} between the multipole range $220 \ge \ell \ge 10$  
EE and BB power spectra obtained from our cleaned map has less power than the corresponding foreground cleaned
spectra obtained from all four WMAP foreground cleaned maps. Power 
spectra for the multipole range $300 \le \ell  \le 600$ are shown in (1,4) and (2,4) figures of Fig. \ref{CompCls}. 
From these plots  we see that, at high $\ell$ 
clean EE and BB spectrum obtained from {\tt QUGMap} have somewhat more power than the WMAP cleaned 
maps.

\section{Validation using Monte-Carlo simulations}
\label{MonteCarlo}
In this section we study our foreground minimization method  for Stokes Q and U parameters 
by performing Monte Carlo simulations. We perform these simulations using CMB, foregrounds 
and detector noise levels compatible to WMAP nine year polarization observations. Before we 
discuss the results of Monte Carlo simulations  we first describe below method of  generating 
input CMB, foreground and detector noise maps. 

\subsection{Input maps} 
\label{valid:model}
\subsubsection{CMB Maps} 
We generate a set of $400$ CMB Stokes Q and U polarization maps at HEALPix pixel resolution parameter 
$N_{\textrm side} = 512$ using a theoretical power spectrum that is consistent with 
Planck 2015 best fit cosmological parameters~\citep{PlanckCosmoParam2016}. We smooth 
each of these maps by a polarized Gaussian beam function of FWHM = $1^{\circ}$.

\subsubsection{Foreground Maps}
To generate Stokes Q and U maps at the frequencies of WMAP K1, KA1, Q, V and W bands we use 
the WMAP nine year base model (that is model `c') for foregrounds as described by ~\cite{Gold2009, Gold2011}.   
The foreground model consists of a synchrotron and a thermal dust polarized component with an 
amplitude and a spectral index defined at each pixel. Specifically, the synchrotron amplitude 
map is provided at the K1 band, which is the frequency band where Synchrotron component is 
most dominant in the WMAP frequency window. The  thermal dust map is provided at the highest
WMAP frequency band W, where the thermal dust component is strongest. The amplitude of Stokes Q 
and U at any given pixel are different while the spectral index is same for both Stokes 
parameter for any given pixel. Since the amplitude maps are given at $N_{\text side } = 64$, 
in antenna Milli Kelvin temperature units,  and  they are already at beam resolution of 
$1^{\circ}$ we simply upgrade their pixel resolution and convert them to thermodynamic 
$\mu K$ temperature unit. The spectral index maps for synchrotron and thermal dust components 
were simply upgraded to $N_{\text side} = 512$. Let $A_s(p)$ ($A_d(p)$) represents the resulting 
synchrotron  (thermal dust)  Q or U map at the reference frequency, $\nu_s =  23$ GHz ($\nu_d = 94$ GHz 
for thermal dust), then, we estimate synchrotron emission in any of the two Stokes Q or U parameter 
and at any WMAP frequency, $\nu$ following
\begin{eqnarray}
A_s(\nu, p) = A_s(p) \left(\frac{\nu}{\nu_s}\right)^{\beta_s(p)}\,,
\label{Synch}
\end{eqnarray}    
where $\beta_s(p)$ is the synchrotron spectral index maps at $N_{\text side} = 512$. In a 
similar fashion, for thermal dust emission, Stokes Q or U parameter map at any of the 
five WMAP frequencies is given by, 
\begin{eqnarray}
A_d(\nu, p) = A_d(p) \left(\frac{\nu}{\nu_d}\right)^{\beta_d(p)}\,.
\label{thdust}
\end{eqnarray} 
We estimate net foreground signal at any frequency $\nu$ for any of the two Stokes parameters Q and U 
by superposing Eqns. \ref{Synch} and \ref{thdust}.  

\subsubsection{Detector Noise Maps}
\begin{figure*}
\includegraphics[scale=0.7]{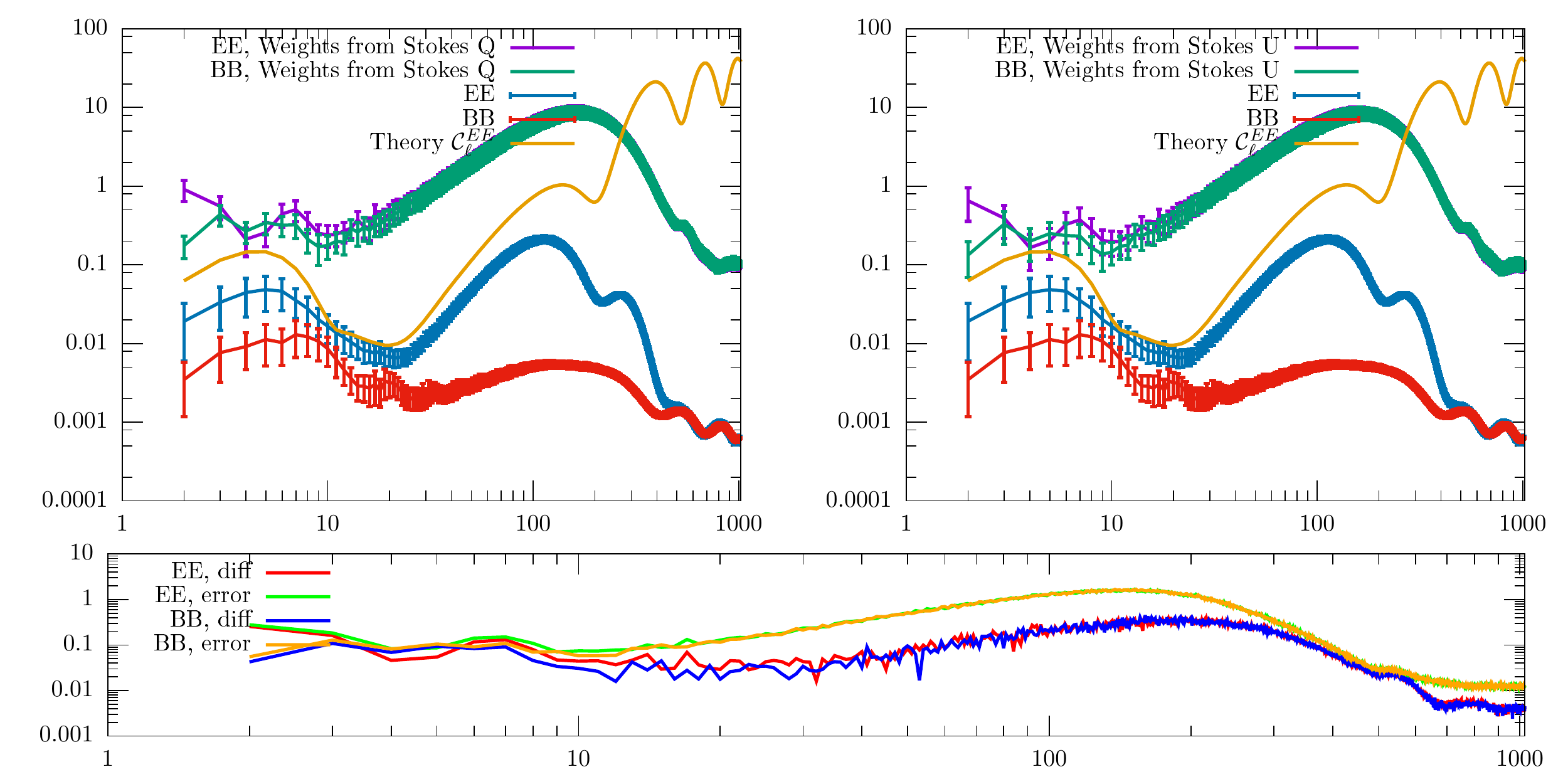}
\caption{In top left panel, mean EE and BB angular power spectra, $\ell(\ell+1)C_{\ell}/2\pi$, in $\mu K^2$
thermodynamic temperature unit, from {\tt CMask} sky region plotted 
in violet and green color. The mean spectra are computed from $289$ Monte Carlo simulations of the  
foreground minimization using the weights estimated from the Stokes Q maps, alone. Violet and green 
curves of top right panel  are same as those of left panel except that these mean spectra are computed 
from $256$ Monte Carlo simulations with weights obtained from Stokes U maps alone. In both these 
panels we show the full sky theoretical EE spectrum expected from LCDM model. The dark blue and red  
curves of both the top  panels represent EE and BB spectra respectively, from  the {\tt CMask} sky, 
from $400$ Monte Carlo simulations of pure CMB Stokes Q and U maps. The error bars represent cosmic
plus sample variance. The red curve, representing the non-zero BB spectrum is actually leakage of 
EE signal to BB on the partial sky. The bottom panel represent the difference of EE spectra (in red) of   
top two panels and the difference of  BB spectra (blue)  of the same  two panels, in $\mu K^2$ thermodynamic 
temperature unit. The green and  
orange lines show $1 \sigma$ error lines  on EE and BB spectra obtained from simulations where weights 
from Stokes Q maps alone are used.     
 }
\label{sim_cls}
\end{figure*}

WMAP science team has provided effective number of observations, $N^i_{obs}(p)$ for each pixel, $p$ for 
a DA, labeled by index $i$ (i = 1,2, ....,10)   as an extension to the corresponding binary fits file
that contain temperature and  polarization anisotropies maps. The standard deviation of noise at each of 
the 10 DA maps per unit observation is mentioned in Table 5 of \cite{Bennett2013} in Milli Kelvin
thermodynamic temperature unit. We convert these standard deviation values in $\mu K$ (thermodynamic) 
temperature unit and  call them $\sigma^i_0$ for the ith DA map.  Using these inputs we estimate noise 
realization for the ith DA map following 
\begin{eqnarray}
n^i(p) = \frac{\sigma^i_0}{\sqrt{N^i_{obs}(p)}}G^i(p)\,,
\end{eqnarray}
where $G^i(p)$ represents a Gaussian deviate with zero mean and unit variance corresponding to the DA 
under consideration. The resulting noise map, $n^i(p)$ is smoothed in the harmonic space 
by the ratio of beam window function corresponding to $1^{\circ}$ polarized Gaussian beam and 
the WMAP nine year supplied beam functions corresponding to different DAs. We generate $400$ noise 
simulations  for each WMAP DAs. We form `frequency-band' noise maps for each of Q, V and W bands by 
averaging over noise maps of all DAs available within a given frequency band. At the end of these 
operations we obtain $400$ noise simulations for each of five WMAP frequency   bands and for 
each of Stokes  Q and U parameters at $1^{\circ}$ beam resolution and at $N_{\textrm side} = 512$.
 We note that noise maps for different frequency maps and different simulations have uncorrelated 
noise properties.    

\subsection{Results}
\label{sim_results}

\begin{figure*}
\includegraphics[scale=0.7]{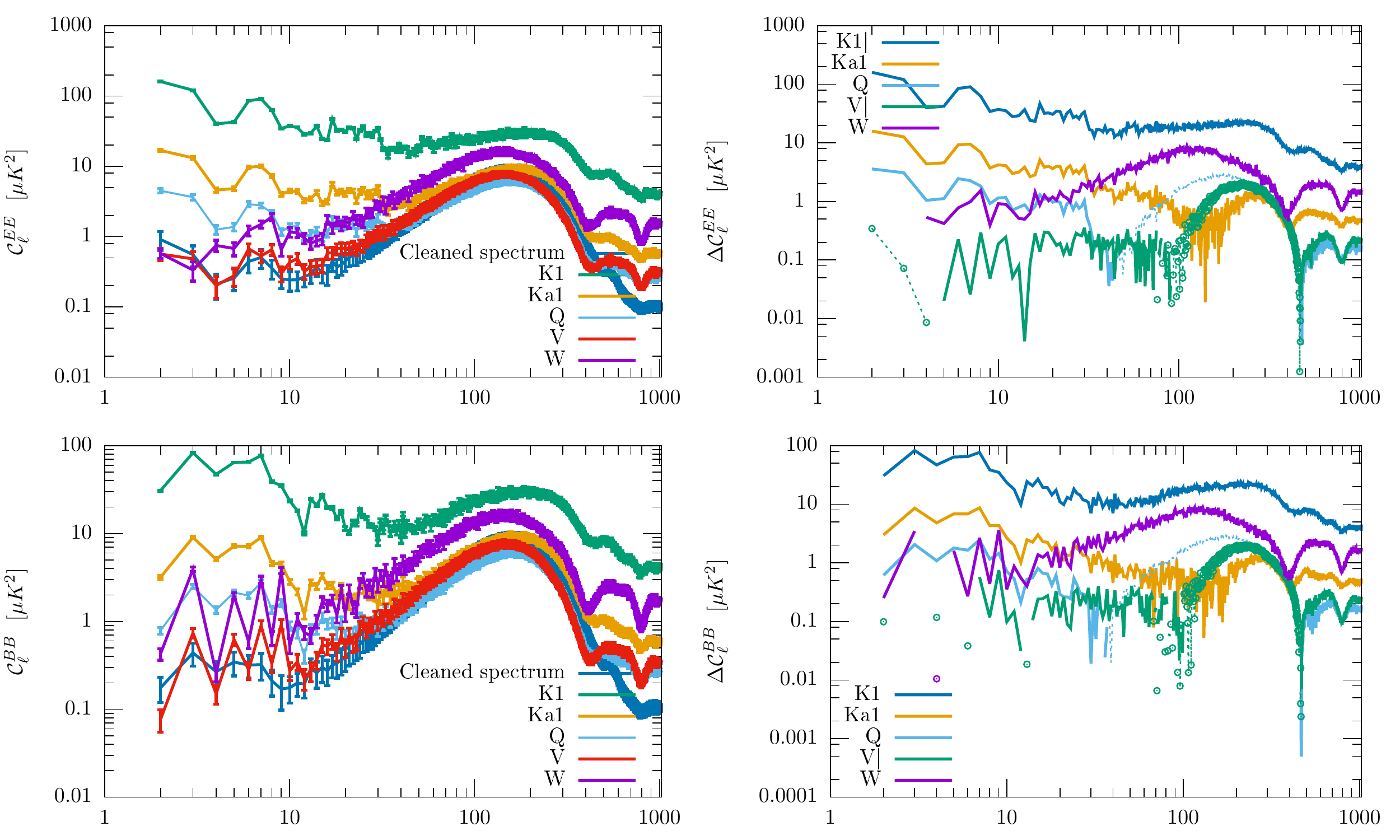}
\caption{Top left and bottom left: figures showing the  mean EE and BB power spectra respectively obtained 
from {\tt CMask} region from cleaned Stokes Q and U maps obtained from Monte Carlo simulations  along with 
the corresponding mean spectra from simulated  input maps  from the 
 {\tt CMask} sky region.  The vertical axes of both the figures show the `reduced spectrum', 
$\ell(\ell+1) C_{\ell}/2\pi$. Top right and bottom right: 
The absolute difference of reduced EE and BB spectra respectively, estimated from our \{Q, U\} cleaned maps 
and five input maps from {\tt CMask} region.  See Section \ref{sim_results} for details.}
\label{inp_out_cls_sim}
\end{figure*}

Using foreground maps corresponding to five WMAP frequency bands, $400$ CMB simulations and additionally 
$400$  noise simulations for each of WMAP frequency maps, as discussed in Section \ref{valid:model} 
we form a total of $400$  realizations  of foreground and detector noise contaminated CMB polarization 
maps for each of Stokes Q and U parameters for each of five WMAP frequencies. This effectively give 
us a set of $400$ realizations, each of which contain five maps corresponding to five WMAP   frequencies 
for each Stokes parameter. As is the case for our analysis on the WMAP data we apply {\tt CMask} on each 
of these maps before foreground minimization.  For the simulation of foreground removal procedure we
separately clean the Q polarization and U polarization maps. For Q polarization maps we minimize the 
kurtosis with respect to weights of three WMAP bands, K1, Q and W respectively. For this purpose we initialize 
the weights in {\tt PolG}  code for K1, Q,  W bands as $-0.3, 0.3, -0.3$ respectively.  We fix the Ka1
band weights to $0.5$ for all the simulations of foreground minimization. For each of these simulations 
V band weight is obtained from the constrained equation that weights must add to unity to preserve the 
CMB Q polarization anisotropy, due to blackbody nature of CMB frequency spectrum. Since CMB polarization
anisotropy is a weak signal compared to detector noise contamination at WMAP frequencies, weights in some 
of these simulations that have not been properly converged  give rise to undesired and high foreground 
residuals. To avoid such simulations from our analysis, from all the $400$ sets 
of weights obtained after the end of all simulations we take only those sets of weights for which the V 
band weight lies between $0.2$ and $1.1$. We find that such a choice of prior on the V band weight 
is sufficient to remove simulations that resulted in any visibly excess power in the  final \{Q, U\} 
cleaned maps.
Using the set of all weights from $289$ simulations for which the V band weights lie  within the chosen 
range of prior, we obtain the foreground cleaned map for Q Stokes parameter from 
the {\tt CMask} region.  We also use the same sets of weights to obtain $289$ foreground cleaned maps 
of U Stokes polarization from the sky region survived after application of  the {\tt CMask}. From each of 
these cleaned partial sky \{Q, U\}  polarization maps we obtain both EE and BB  angular power spectra. 
We show the average of all such EE and BB  spectra, $\ell(\ell+1)C_{\ell}/2\pi$, in $\mu K^2$ thermodynamic 
temperature unit,  along with the error-bars estimated from these 
simulations in top left panel of Fig. \ref{sim_cls}. We note that the non-zero BB spectrum in this plot indicates 
presence of residual foregrounds and detector noise, and does not represent any  primordial BB 
power spectrum, since the initial theoretical power spectrum from which the CMB Stokes Q and U 
polarization maps were generated contained no BB spectrum. Some portion of the BB spectrum seen in 
this figure is caused due to leakage of EE signal to BB signal due finite sky area effect.        

\begin{figure*}
\includegraphics[scale=0.7]{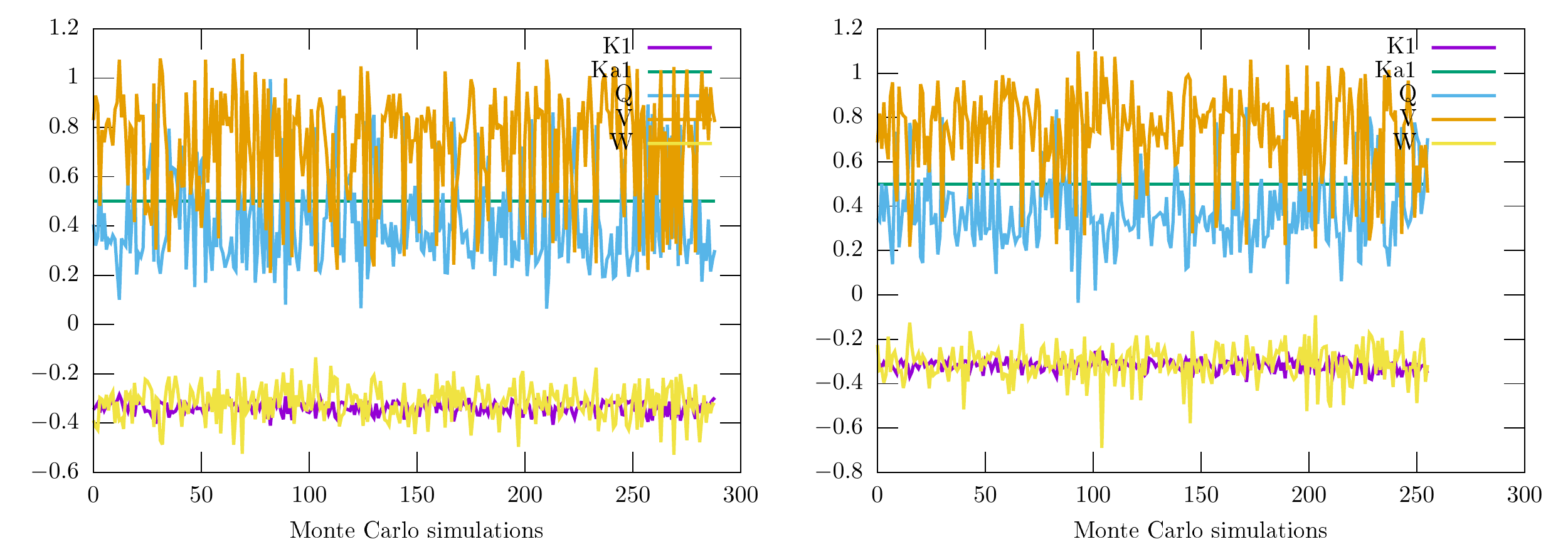}
\caption{Variation of weights for different frequency maps for different Monte Carlo simulations of 
our foreground removal procedure. Left panel represents the case when weights are estimated from the 
Q Stokes maps alone. The right panel shows the variation when weights are estimated from the Stokes 
U maps alone.}
\label{weight_plot}
\end{figure*}

Following a  procedure similar to estimating the partial sky foreground cleaned \{Q, U\} maps using the weights 
obtained from Q Stokes maps alone, we also estimate a set of cleaned maps \{Q, U\} maps from the same sky region 
using weights that are estimated using the U maps alone.  In this case, we initialize, K1, Q and W  band 
weights to the identical values as the Q polarization case discussed in the preceding paragraph. We fix 
the Ka1  band weight to $0.5$ and the {\tt PolG} code then estimates the V band weights from the constrained equation 
satisfied by the weights. Imposing the prior $[0.2, 1.1]$ on the resulting $400$  V band weights we find that 
a set of $256$ simulations for which the final V band weights fulfill the prior condition. Using these weights 
we estimate foreground cleaned \{Q, U\} maps from the sky region that survives after application of {\tt CMask}. 
We estimate partial sky EE and BB angular power spectra from each of these foreground cleaned maps. In top right 
panel of Fig. \ref{sim_cls}  we show the mean of all such spectra along with the error-bar computed from the 
same set of simulations. The mean EE and BB spectra match well with the corresponding 
spectra obtained from the cleaned maps that were estimated from the weights obtained from Q polarization maps 
alone. The difference of EE spectra of top  panels is shown in red line in bottom panel. Also shown in bottom 
panel of Fig. \ref{sim_cls}, the difference 
of BB spectra  in blue.  The green and blue curves of the bottom panel shows $1\sigma$  error lines. From the difference 
spectra and error line plots we conclude that our foreground minimization method works overall with the similar efficiency for both Stokes Q and U 
polarizations maps separately. We again emphasize that the BB spectra in this case is mere indication of presence 
of residual foreground and detector noise in the final cleaned maps, and not the primordial BB signal.

We compare the mean cleaned EE and BB power spectra  estimated from accepted  Monte Carlo simulations (for the case 
when weights are determined by the Q Stokes maps) along  with  the mean EE and BB spectra of  each of five simulated
frequency maps from the $400$ simulations in top left and bottom left of  Fig. \ref{inp_out_cls_sim}. All spectra of 
these figures are estimated from {\tt CMask} sky region and are multiplied by the factor $\ell(\ell+1) /2\pi$.  
The error bars of these figures are estimated from Monte Carlo simulations. The reduced power in the 
cleaned maps indicate efficiency of foreground minimization using the method described in this work. 
The top right figure represent the absolute difference of mean cleaned EE spectrum and  the mean spectra of 
each of five input maps of Monte Carlo simulations and represent the appropriate difference spectra using the 
results shown in top left panel.  The solid lines of top right panel  represent the 
case when the mean EE spectrum has less power than the power of an input map, implying an effective foreground removal 
has taken place for the particular frequency band and for the particular multipole(s). The dashed lines and points 
of these plots represent multipoles for which mean cleaned EE spectrum is higher than the spectrum of an input 
frequency map. Hence foreground minimization at these multipoles and frequency band is not very effective. The 
bottom right figure of Fig. \ref{inp_out_cls_sim} shows the same as top right figure, except that the bottom 
right figure shows the case for BB spectra.

How does the weights vary with the simulations? We have shown the weights for each of K1, Ka1, Q, V 
and W bands from the accepted simulations as a function of simulation index in Fig. \ref{weight_plot}. 
Left panel of this figure shows the case when weights  are estimated only from Q Stokes maps alone. The right
panel shows the case when weights are estimated from the U polarization maps alone. In both cases we see 
that weights for K1 band have minimum dispersion since it has the maximum signal to noise ratio for the 
foregrounds due to strong synchrotron emission. The V band weights have  the largest dispersion since 
V band weights are estimated passively, i.e., from the constrained equation of weights given the values 
of weights of rest of the frequency bands.  

\section{Discussion and Conclusion} 
\label{D&C}
Simple models of inflation predict that the fluctuations in the primordial inflaton
field were Gaussian to very good approximations. The Gaussian nature of the primordial 
density fluctuations directly imply that the CMB anisotropies also follow Gaussian
distribution, since the later is related to the former by means of a 
multiplicative transfer function. Using the concept of Gaussian distribution 
of CMB polarization anisotropies and strongly non-Gaussian nature of diffused polarized 
emissions from Milky Way, in this work we have proposed a methodology to  separate 
the former from the observed mixture of the later using the WMAP nine year Stokes Q
and U polarization maps, from a region of the sky that excludes regions heavily contaminated 
by foregrounds or  detector noise.  

For cosmological analysis of observed CMB polarization maps a major task is to 
minimize the foreground contaminations. The foreground polarization may be rather
complex -- apart from variation of spectral spectral 
indices with sky locations for both synchrotron and thermal dust Stokes Q and U 
signals, thermal dust emission may exhibit high fraction of polarization even from 
regions where the signal is weak. In this context, the reconstruction of CMB polarization maps 
presented in this article has two important advantages. First, since the method uses 
usual ILC technique as the basic driving machinery, one  does not need to use any 
explicit model for polarized foregrounds. Thus our method possesses an interesting 
property that the foreground minimized CMB Stokes maps are free from any error that
could otherwise result from any incorrect foreground modeling. The only  
assumption we make about the foreground spectrum is that they are non-blackbody. 
This is a valid assumption for polarized emissions like synchrotron and thermal 
dust. The second, and very important in the context of physics of early 
universe, advantage of  our method, is that, our method minimizes a measure of 
non-Gaussian properties to estimate the foreground minimized CMB polarization map. The 
assumption of Gaussian properties of CMB anisotropies is motivated from the 
simplest models of inflation. The non-Gaussian nature of polarized foreground 
emission is expected due to various non-linear mechanisms that operate during 
their emission. We demonstrate strong non-Gaussian properties of polarized 
foregrounds based upon empirical results. 

We have critically studied the performance of our polarized foreground minimization. Based upon these 
studies we conclude that the method removes significant amount of foregrounds from different frequency bands. 
The performance of the method  is expected to be even superior for future CMB polarization experiments with low level of detector noise
compared to the primordial CMB polarization signal, 
since detector noise plays a stringent `bottle-neck' for effective foreground minimization. Interestingly, our 
foreground minimized Stoke Q and U maps have less power (both EE and BB) 
than the WMAP template cleaned maps. This shows that,  when analyzed over the 
{\tt CMask} region our method performs a better foreground removal than a global
foreground removal method as followed by the WMAP Science team using template 
models for the polarized foregrounds. In spite of this, we make a cautious comment that 
there is  indeed residual foregrounds and detector noise  in our cleaned Stokes Q and U maps. 
Great care must be taken, for cosmological analysis of these cleaned maps.  It would be very useful to apply our method 
on the polarization observations of Planck and polarization specific future CMB 
missions. 

Analyzing our cleaned EE and BB spectra we find that performance of foreground removal 
depends upon the multipoles. This is expected since foreground spectra themselves depend
upon multipoles and moreover detector noise usually become stronger at higher multipoles. 
To achieve a better foreground removal an  interesting future project will be to generalize 
our method in the multipole space. 

One interesting property of our method is that, although it minimizes non-Gaussian
property of foregrounds, any non-Gaussianities intrinsic to CMB  remains preserved 
in the final cleaned map, since CMB follows blackbody spectrum with high accuracy. 
The condition that weights for all frequency maps add to unity preserves CMB polarization
signal even if it has some minute level of non-Gaussianities. In this context we 
mention that our method will be able to provide a complete picture of CMB 
anisotropies for models of inflation that produces some level of primordial 
non-Gaussianity, e.g., see \cite{Wands2010} for a discussion of local non-Gaussian 
from inflation, or \cite{Bartolo2004} and \cite{PlanckNG2016, PlanckNG2014} with references therein for  comprehensive reviews
about non-Gaussianities generated from different models of inflation. It will be valuable to use our method on the future generation
CMB experiments to constrain such models that predict minute levels of 
primordial non-Gaussianities.
   
 We use publicly available HEALPix~\cite{Gorski2005} package available from 
http://healpix.sourceforge.net to perform forward and backward spherical harmonic transformations and 
for visualization purposes. We acknowledge the use of the Legacy Archive for Microwave Background 
Data Analysis (LAMBDA). LAMBDA is a part of the High Energy Astrophysics Science Archive Center (HEASARC). 
HEASARC/LAMBDA is supported by the Astrophysics Science Division at the NASA Goddard Space Flight
Center.





\begin{thebibliography}{}
\expandafter\ifx\csname natexlab\endcsname\relax\def\natexlab#1{#1}\fi

\bibitem[{{Acquaviva} {et~al.}(2003){Acquaviva}, {Bartolo}, {Matarrese}, \&
  {Riotto}}]{Acquaviva2003}
{Acquaviva}, V., {Bartolo}, N., {Matarrese}, S., \& {Riotto}, A. 2003, Nuclear
  Physics B, 667, 119

\bibitem[{{Allen} {et~al.}(1987){Allen}, {Grinstein}, \& {Wise}}]{Allen1987}
{Allen}, T.~J., {Grinstein}, B., \& {Wise}, M.~B. 1987, Physics Letters B, 197,
  66

\bibitem[{{Baccigalupi} {et~al.}(2004){Baccigalupi}, {Perrotta}, {de Zotti},
  {Smoot}, {Burigana}, {Maino}, {Bedini}, \& {Salerno}}]{ICA_2004}
{Baccigalupi}, C., {Perrotta}, F., {de Zotti}, G., {et~al.} 2004, \mnras, 354,
  55

\bibitem[{{Balaji} {et~al.}(2003){Balaji}, {Brandenberger}, \&
  {Easson}}]{Balaji2003}
{Balaji}, K.~R.~S., {Brandenberger}, R.~H., \& {Easson}, D.~A. 2003, \jcap, 12,
  008

\bibitem[{{Bartolo} {et~al.}(2004){Bartolo}, {Komatsu}, {Matarrese}, \&
  {Riotto}}]{Bartolo2004}
{Bartolo}, N., {Komatsu}, E., {Matarrese}, S., \& {Riotto}, A. 2004, \physrep,
  402, 103

\bibitem[{{Basak} \& {Delabrouille}(2012)}]{Basak2012}
{Basak}, S., \& {Delabrouille}, J. 2012, \mnras, 419, 1163

\bibitem[{{Basak} \& {Delabrouille}(2013)}]{Basak2013}
---. 2013, \mnras, 435, 18

\bibitem[{{Bennett} {et~al.}(1992){Bennett}, {Smoot}, {Hinshaw}, {Wright},
  {Kogut}, {de Amici}, {Meyer}, {Weiss}, {Wilkinson}, {Gulkis}, {Janssen},
  {Boggess}, {Cheng}, {Hauser}, {Kelsall}, {Mather}, {Moseley}, {Murdock}, \&
  {Silverberg}}]{Bennett1992}
{Bennett}, C.~L., {Smoot}, G.~F., {Hinshaw}, G., {et~al.} 1992, \apjl, 396, L7

\bibitem[{{Bennett} {et~al.}(2003){Bennett}, {Hill}, {Hinshaw}, {Nolta},
  {Odegard}, {Page}, {Spergel}, {Weiland}, {Wright}, {Halpern}, {Jarosik},
  {Kogut}, {Limon}, {Meyer}, {Tucker}, \& {Wollack}}]{Bennett2003}
{Bennett}, C.~L., {Hill}, R.~S., {Hinshaw}, G., {et~al.} 2003, \apjs, 148, 97

\bibitem[{{Bennett} {et~al.}(2013){Bennett}, {Larson}, {Weiland}, {Jarosik},
  {Hinshaw}, {Odegard}, {Smith}, {Hill}, {Gold}, {Halpern}, {Komatsu}, {Nolta},
  {Page}, {Spergel}, {Wollack}, {Dunkley}, {Kogut}, {Limon}, {Meyer}, {Tucker},
  \& {Wright}}]{Bennett2013}
{Bennett}, C.~L., {Larson}, D., {Weiland}, J.~L., {et~al.} 2013, \apjs, 208, 20

\bibitem[{{Bouchet} {et~al.}(1999){Bouchet}, {Prunet}, \&
  {Sethi}}]{Bouchet1999}
{Bouchet}, F.~R., {Prunet}, S., \& {Sethi}, S.~K. 1999, \mnras, 302, 663

\bibitem[{{Bucher}(2015)}]{Bucher2015}
{Bucher}, M. 2015, International Journal of Modern Physics D, 24, 1530004

\bibitem[{{Bucher} {et~al.}(2001){Bucher}, {Moodley}, \& {Turok}}]{Bucher2001}
{Bucher}, M., {Moodley}, K., \& {Turok}, N. 2001, Physical Review Letters, 87,
  191301

\bibitem[{{Bunn} {et~al.}(1994){Bunn}, {Fisher}, {Hoffman}, {Lahav}, {Silk}, \&
  {Zaroubi}}]{Bunn1994}
{Bunn}, E.~F., {Fisher}, K.~B., {Hoffman}, Y., {et~al.} 1994, \apjl, 432, L75

\bibitem[{{Cabella} \& {Kamionkowski}(2004)}]{Cabella2004}
{Cabella}, P., \& {Kamionkowski}, M. 2004, ArXiv Astrophysics e-prints,
  astro-ph/0403392

\bibitem[{{Crittenden} {et~al.}(1993){Crittenden}, {Davis}, \&
  {Steinhardt}}]{CDS_1993}
{Crittenden}, R., {Davis}, R.~L., \& {Steinhardt}, P.~J. 1993, \apjl, 417, L13

\bibitem[{{Crittenden} {et~al.}(1995){Crittenden}, {Coulson}, \&
  {Turok}}]{cct_1995}
{Crittenden}, R.~G., {Coulson}, D., \& {Turok}, N.~G. 1995, \prd, 52, R5402

\bibitem[{{Delabrouille} {et~al.}(2009){Delabrouille}, {Cardoso}, {Le Jeune},
  {Betoule}, {Fay}, \& {Guilloux}}]{Delabrouille2009}
{Delabrouille}, J., {Cardoso}, J.-F., {Le Jeune}, M., {et~al.} 2009, \aap, 493,
  835

\bibitem[{{Eriksen} {et~al.}(2004){Eriksen}, {Banday}, {G{\'o}rski}, \&
  {Lilje}}]{LILC_Eriksen04}
{Eriksen}, H.~K., {Banday}, A.~J., {G{\'o}rski}, K.~M., \& {Lilje}, P.~B. 2004,
  \apj, 612, 633

\bibitem[{{Eriksen} {et~al.}(2008{\natexlab{a}}){Eriksen}, {Dickinson},
  {Jewell}, {Banday}, {G{\'o}rski}, \& {Lawrence}}]{Eriksen2008a}
{Eriksen}, H.~K., {Dickinson}, C., {Jewell}, J.~B., {et~al.}
  2008{\natexlab{a}}, \apjl, 672, L87

\bibitem[{{Eriksen} {et~al.}(2008{\natexlab{b}}){Eriksen}, {Jewell},
  {Dickinson}, {Banday}, {G{\'o}rski}, \& {Lawrence}}]{Eriksen2008}
{Eriksen}, H.~K., {Jewell}, J.~B., {Dickinson}, C., {et~al.}
  2008{\natexlab{b}}, \apj, 676, 10

\bibitem[{{Eriksen} {et~al.}(2007){Eriksen}, {Huey}, {Saha}, {Hansen}, {Dick},
  {Banday}, {G{\'o}rski}, {Jain}, {Jewell}, {Knox}, {Larson}, {O'Dwyer},
  {Souradeep}, \& {Wandelt}}]{Eriksen2007}
{Eriksen}, H.~K., {Huey}, G., {Saha}, R., {et~al.} 2007, \apj, 656, 641

\bibitem[{{Finkbeiner} {et~al.}(1999){Finkbeiner}, {Davis}, \&
  {Schlegel}}]{Fink1999}
{Finkbeiner}, D.~P., {Davis}, M., \& {Schlegel}, D.~J. 1999, \apj, 524, 867

\bibitem[{{Fixsen} {et~al.}(1996){Fixsen}, {Cheng}, {Gales}, {Mather},
  {Shafer}, \& {Wright}}]{Fixen1996}
{Fixsen}, D.~J., {Cheng}, E.~S., {Gales}, J.~M., {et~al.} 1996, \apj, 473, 576

\bibitem[{{Frewin} {et~al.}(1994){Frewin}, {Polnarev}, \& {Coles}}]{Frewin1994}
{Frewin}, R.~A., {Polnarev}, A.~G., \& {Coles}, P. 1994, \mnras, 266, L21

\bibitem[{{Gangui} {et~al.}(1994){Gangui}, {Lucchin}, {Matarrese}, \&
  {Mollerach}}]{Gangui1994}
{Gangui}, A., {Lucchin}, F., {Matarrese}, S., \& {Mollerach}, S. 1994, \apj,
  430, 447

\bibitem[{{Gangui} {et~al.}(2002){Gangui}, {Martin}, \&
  {Sakellariadou}}]{Gangui2002}
{Gangui}, A., {Martin}, J., \& {Sakellariadou}, M. 2002, \prd, 66, 083502

\bibitem[{{Gold} {et~al.}(2009){Gold}, {Bennett}, {Hill}, {Hinshaw}, {Odegard},
  {Page}, {Spergel}, {Weiland}, {Dunkley}, {Halpern}, {Jarosik}, {Kogut},
  {Komatsu}, {Larson}, {Meyer}, {Nolta}, {Wollack}, \& {Wright}}]{Gold2009}
{Gold}, B., {Bennett}, C.~L., {Hill}, R.~S., {et~al.} 2009, \apjs, 180, 265

\bibitem[{{Gold} {et~al.}(2011){Gold}, {Odegard}, {Weiland}, {Hill}, {Kogut},
  {Bennett}, {Hinshaw}, {Chen}, {Dunkley}, {Halpern}, {Jarosik}, {Komatsu},
  {Larson}, {Limon}, {Meyer}, {Nolta}, {Page}, {Smith}, {Spergel}, {Tucker},
  {Wollack}, \& {Wright}}]{Gold2011}
{Gold}, B., {Odegard}, N., {Weiland}, J.~L., {et~al.} 2011, \apjs, 192, 15

\bibitem[{{G{\'o}rski} {et~al.}(2005){G{\'o}rski}, {Hivon}, {Banday},
  {Wandelt}, {Hansen}, {Reinecke}, \& {Bartelmann}}]{Gorski2005}
{G{\'o}rski}, K.~M., {Hivon}, E., {Banday}, A.~J., {et~al.} 2005, \apj, 622,
  759

\bibitem[{{Guth}(1981)}]{GuthInflation1981}
{Guth}, A.~H. 1981, \prd, 23, 347

\bibitem[{{Hansen} \& {G{\'o}rski}(2003)}]{Gabor2003}
{Hansen}, F.~K., \& {G{\'o}rski}, K.~M. 2003, \mnras, 343, 559

\bibitem[{{Hinshaw} {et~al.}(2007){Hinshaw}, {Nolta}, {Bennett}, {Bean},
  {Dore}, {Greason}, {Halpern}, {Hill}, {Jarosik}, {Kogut}, {Komatsu}, {Limon},
  {Odegard}, {Meyer}, {Page}, {Peiris}, {Spergel}, {Tucker}, {Verde},
  {Weiland}, {Wollack}, \& {Wright}}]{Hinshaw_07}
{Hinshaw}, G., {Nolta}, M.~R., {Bennett}, C.~L., {et~al.} 2007, Astrophys. J.
  Suppl. Ser., 170, 288

\bibitem[{{Hu} \& {Okamoto}(2002)}]{Hu2002}
{Hu}, W., \& {Okamoto}, T. 2002, \apj, 574, 566

\bibitem[{Hyv\"{a}rinen \& Oja(2000)}]{ICA_Hyvarinen}
Hyv\"{a}rinen, A., \& Oja, E. 2000, Neural Netw., 13, 411

\bibitem[{{Kamionkowski} {et~al.}(1997){Kamionkowski}, {Kosowsky}, \&
  {Stebbins}}]{KKS1997}
{Kamionkowski}, M., {Kosowsky}, A., \& {Stebbins}, A. 1997, Physical Review
  Letters, 78, 2058

\bibitem[{Kim {et~al.}(2008)Kim, Naselsky, \& Christensen}]{Kim_HILC_temp}
Kim, J., Naselsky, P., \& Christensen, P.~R. 2008, Phys. Rev. D, 77, 103002

\bibitem[{Kim {et~al.}(2009)Kim, Naselsky, \& Christensen}]{Kim_HILC_Pol}
---. 2009, Phys. Rev. D, 79, 023003

\bibitem[{{Knox} \& {Song}(2002)}]{Knox2002}
{Knox}, L., \& {Song}, Y.-S. 2002, Physical Review Letters, 89, 011303

\bibitem[{{Komatsu} {et~al.}(2003){Komatsu}, {Kogut}, {Nolta}, {Bennett},
  {Halpern}, {Hinshaw}, {Jarosik}, {Limon}, {Meyer}, {Page}, {Spergel},
  {Tucker}, {Verde}, {Wollack}, \& {Wright}}]{Komatsu2003}
{Komatsu}, E., {Kogut}, A., {Nolta}, M.~R., {et~al.} 2003, \apjs, 148, 119

\bibitem[{{Kosowsky}(1996)}]{Kosowsky1996}
{Kosowsky}, A. 1996, Annals of Physics, 246, 49

\bibitem[{{Kosowsky}(1999)}]{Kosowsky1999}
---. 1999, \nar, 43, 157

\bibitem[{{Lewis} {et~al.}(2002){Lewis}, {Challinor}, \& {Turok}}]{Lewis2002}
{Lewis}, A., {Challinor}, A., \& {Turok}, N. 2002, \prd, 65, 023505

\bibitem[{{Linde}(1983)}]{LindeInflation1983}
{Linde}, A.~D. 1983, Physics Letters B, 129, 177

\bibitem[{{Lubin} {et~al.}(1983){Lubin}, {Melese}, \& {Smoot}}]{Lubin1983}
{Lubin}, P., {Melese}, P., \& {Smoot}, G. 1983, \apjl, 273, L51

\bibitem[{{Maldacena}(2003)}]{Maldacena2003}
{Maldacena}, J. 2003, Journal of High Energy Physics, 5, 013

\bibitem[{{Mather} {et~al.}(1990){Mather}, {Cheng}, {Eplee}, {Isaacman},
  {Meyer}, {Shafer}, {Weiss}, {Wright}, {Bennett}, {Boggess}, {Dwek}, {Gulkis},
  {Hauser}, {Janssen}, {Kelsall}, {Lubin}, {Moseley}, {Murdock}, {Silverberg},
  {Smoot}, \& {Wilkinson}}]{Mather1990}
{Mather}, J.~C., {Cheng}, E.~S., {Eplee}, Jr., R.~E., {et~al.} 1990, \apjl,
  354, L37

\bibitem[{{Mather} {et~al.}(1994){Mather}, {Cheng}, {Cottingham}, {Eplee},
  {Fixsen}, {Hewagama}, {Isaacman}, {Jensen}, {Meyer}, {Noerdlinger}, {Read},
  {Rosen}, {Shafer}, {Wright}, {Bennett}, {Boggess}, {Hauser}, {Kelsall},
  {Moseley}, {Silverberg}, {Smoot}, {Weiss}, \& {Wilkinson}}]{Mather1994}
{Mather}, J.~C., {Cheng}, E.~S., {Cottingham}, D.~A., {et~al.} 1994, \apj, 420,
  439

\bibitem[{{Munshi} {et~al.}(1995){Munshi}, {Souradeep}, \&
  {Starobinsky}}]{Munshi1995}
{Munshi}, D., {Souradeep}, T., \& {Starobinsky}, A.~A. 1995, \apj, 454, 552

\bibitem[{{Negroponte} \& {Silk}(1980)}]{Negroponte1980}
{Negroponte}, J., \& {Silk}, J. 1980, Physical Review Letters, 44, 1433

\bibitem[{{Padmanabhan} \& {Finkbeiner}(2005)}]{NikhilDoug2005}
{Padmanabhan}, N., \& {Finkbeiner}, D.~P. 2005, \prd, 72, 023508

\bibitem[{{Page} {et~al.}(2007){Page}, {Hinshaw}, {Komatsu}, {Nolta},
  {Spergel}, {Bennett}, {Barnes}, {Bean}, {Dor{\'e}}, {Dunkley}, {Halpern},
  {Hill}, {Jarosik}, {Kogut}, {Limon}, {Meyer}, {Odegard}, {Peiris}, {Tucker},
  {Verde}, {Weiland}, {Wollack}, \& {Wright}}]{Page2007}
{Page}, L., {Hinshaw}, G., {Komatsu}, E., {et~al.} 2007, \apjs, 170, 335

\bibitem[{{Penzias} \& {Wilson}(1965)}]{Penzias1965}
{Penzias}, A.~A., \& {Wilson}, R.~W. 1965, \apj, 142, 419

\bibitem[{{Planck Collaboration} {et~al.}(2014){Planck Collaboration}, {Ade},
  {Aghanim}, {Armitage-Caplan}, {Arnaud}, {Ashdown}, {Atrio-Barandela},
  {Aumont}, {Baccigalupi}, {Banday}, \& et~al.}]{PlanckNG2014}
{Planck Collaboration}, {Ade}, P.~A.~R., {Aghanim}, N., {et~al.} 2014, \aap,
  571, A24

\bibitem[{{Planck Collaboration} {et~al.}(2016{\natexlab{a}}){Planck
  Collaboration}, {Adam}, {Ade}, {Aghanim}, {Arnaud}, {Ashdown}, {Aumont},
  {Baccigalupi}, {Banday}, {Barreiro}, \& et~al.}]{PlanckCol2016b}
{Planck Collaboration}, {Adam}, R., {Ade}, P.~A.~R., {et~al.}
  2016{\natexlab{a}}, \aap, 594, A9

\bibitem[{{Planck Collaboration} {et~al.}(2016{\natexlab{b}}){Planck
  Collaboration}, {Adam}, {Ade}, {Aghanim}, {Alves}, {Arnaud}, {Ashdown},
  {Aumont}, {Baccigalupi}, {Banday}, \& et~al.}]{PlanckCol2016a}
---. 2016{\natexlab{b}}, \aap, 594, A10

\bibitem[{{Planck Collaboration} {et~al.}(2016{\natexlab{c}}){Planck
  Collaboration}, {Ade}, {Aghanim}, {Arnaud}, {Ashdown}, {Aumont},
  {Baccigalupi}, {Banday}, {Barreiro}, {Bartlett}, \&
  et~al.}]{PlanckCosmoParam2016}
{Planck Collaboration}, {Ade}, P.~A.~R., {Aghanim}, N., {et~al.}
  2016{\natexlab{c}}, \aap, 594, A13

\bibitem[{{Planck Collaboration} {et~al.}(2016{\natexlab{d}}){Planck
  Collaboration}, {Ade}, {Aghanim}, {Arnaud}, {Arroja}, {Ashdown}, {Aumont},
  {Baccigalupi}, {Ballardini}, {Banday}, \& et~al.}]{PlanckNG2016}
---. 2016{\natexlab{d}}, \aap, 594, A17

\bibitem[{{Planck Collaboration} {et~al.}(2016{\natexlab{e}}){Planck
  Collaboration}, {Ade}, {Aghanim}, {Arnaud}, {Arroja}, {Ashdown}, {Aumont},
  {Baccigalupi}, {Ballardini}, {Banday}, \& et~al.}]{PlanckInflation2016}
---. 2016{\natexlab{e}}, \aap, 594, A20

\bibitem[{{Polnarev}(1985)}]{Polnarev1985}
{Polnarev}, A.~G. 1985, \sovast, 29, 607

\bibitem[{{Press} {et~al.}(1989){Press}, {Flannery}, {Teukolsky}, \&
  {Vetterling}}]{nrc}
{Press}, W.~H., {Flannery}, B.~P., {Teukolsky}, S.~A., \& {Vetterling}, W.~T.
  1989, {Numerical recipes in C. The art of scientific computing}

\bibitem[{{Rees}(1968)}]{Rees1968}
{Rees}, M.~J. 1968, \apjl, 153, L1

\bibitem[{{Remazeilles} {et~al.}(2011{\natexlab{a}}){Remazeilles},
  {Delabrouille}, \& {Cardoso}}]{Remazeilles2011a}
{Remazeilles}, M., {Delabrouille}, J., \& {Cardoso}, J.-F. 2011{\natexlab{a}},
  \mnras, 410, 2481

\bibitem[{{Remazeilles} {et~al.}(2011{\natexlab{b}}){Remazeilles},
  {Delabrouille}, \& {Cardoso}}]{Remazeilles2011}
---. 2011{\natexlab{b}}, ArXiv e-prints, arXiv:1103.1166

\bibitem[{{Rogers} {et~al.}(2016{\natexlab{a}}){Rogers}, {Peiris}, {Leistedt},
  {McEwen}, \& {Pontzen}}]{SILC_2016_temp}
{Rogers}, K.~K., {Peiris}, H.~V., {Leistedt}, B., {McEwen}, J.~D., \&
  {Pontzen}, A. 2016{\natexlab{a}}, \mnras, 460, 3014

\bibitem[{{Rogers} {et~al.}(2016{\natexlab{b}}){Rogers}, {Peiris}, {Leistedt},
  {McEwen}, \& {Pontzen}}]{SILC_2016_Pol}
---. 2016{\natexlab{b}}, \mnras, 463, 2310

\bibitem[{{Saha}(2011)}]{Saha2011}
{Saha}, R. 2011, \apjl, 739, L56

\bibitem[{{Saha} \& {Aluri}(2016)}]{Saha2016}
{Saha}, R., \& {Aluri}, P.~K. 2016, \apj, 829, 113

\bibitem[{{Saha} {et~al.}(2006){Saha}, {Jain}, \& {Souradeep}}]{Saha2006}
{Saha}, R., {Jain}, P., \& {Souradeep}, T. 2006, \apjl, 645, L89

\bibitem[{{Saha} {et~al.}(2008){Saha}, {Prunet}, {Jain}, \&
  {Souradeep}}]{Saha2008}
{Saha}, R., {Prunet}, S., {Jain}, P., \& {Souradeep}, T. 2008, \prd, 78, 023003

\bibitem[{{Samal} {et~al.}(2010){Samal}, {Saha}, {Delabrouille}, {Prunet},
  {Jain}, \& {Souradeep}}]{Samal2010}
{Samal}, P.~K., {Saha}, R., {Delabrouille}, J., {et~al.} 2010, \apj, 714, 840

\bibitem[{{Samal} {et~al.}(2008){Samal}, {Saha}, {Jain}, \&
  {Ralston}}]{Samal2008}
{Samal}, P.~K., {Saha}, R., {Jain}, P., \& {Ralston}, J.~P. 2008, \mnras, 385,
  1718

\bibitem[{{Seljak} \& {Zaldarriaga}(1999)}]{SelZal1999}
{Seljak}, U., \& {Zaldarriaga}, M. 1999, Physical Review Letters, 82, 2636

\bibitem[{{Spergel} \& {Zaldarriaga}(1997)}]{SpZal1997}
{Spergel}, D.~N., \& {Zaldarriaga}, M. 1997, Physical Review Letters, 79, 2180

\bibitem[{{Stark}(1981{\natexlab{a}})}]{Stark1981_b}
{Stark}, R.~F. 1981{\natexlab{a}}, \mnras, 195, 127

\bibitem[{{Stark}(1981{\natexlab{b}})}]{Stark1981_a}
---. 1981{\natexlab{b}}, \mnras, 195, 115

\bibitem[{{Starobinsky}(1982)}]{StarobinskyInflation1982}
{Starobinsky}, A.~A. 1982, Physics Letters B, 117, 175

\bibitem[{{Sudevan} {et~al.}(2017){Sudevan}, {Aluri}, {Yadav}, {Saha}, \&
  {Souradeep}}]{Saha2017}
{Sudevan}, V., {Aluri}, P.~K., {Yadav}, S.~K., {Saha}, R., \& {Souradeep}, T.
  2017, \apj, 842, 62

\bibitem[{{Tegmark} \& {de Oliveira-Costa}(2001)}]{Tegmark2001}
{Tegmark}, M., \& {de Oliveira-Costa}, A. 2001, \prd, 64, 063001

\bibitem[{{Tegmark} {et~al.}(2003){Tegmark}, {de Oliveira-Costa}, \&
  {Hamilton}}]{Tegmark2003}
{Tegmark}, M., {de Oliveira-Costa}, A., \& {Hamilton}, A.~J. 2003, Phys. Rev.
  D, 68, 123523

\bibitem[{{Tegmark} \& {Efstathiou}(1996{\natexlab{a}})}]{Tegmark1996}
{Tegmark}, M., \& {Efstathiou}, G. 1996{\natexlab{a}}, \mnras, 281, 1297

\bibitem[{{Tegmark} \& {Efstathiou}(1996{\natexlab{b}})}]{Tegmark96}
---. 1996{\natexlab{b}}, Mon. Not. R. Astron. Soc., 281, 1297

\bibitem[{{Wands}(2010)}]{Wands2010}
{Wands}, D. 2010, Classical and Quantum Gravity, 27, 124002

\bibitem[{{Zaldarriaga}(1997)}]{Zaldarriaga1997}
{Zaldarriaga}, M. 1997, \prd, 55, 1822

\bibitem[{{Zaldarriaga} \& {Seljak}(1998)}]{ZalSel1998}
{Zaldarriaga}, M., \& {Seljak}, U. 1998, \prd, 58, 023003

\end{thebibliography}


\end{document}